\documentclass[traditabstract]{aa} % for the abstract without structuration 
\usepackage{graphicx}
\usepackage{natbib}
\usepackage{bm}
\usepackage{txfonts}
\begin{document}
   \title{The magnetorotational instability as a jet launching mechanism}

   \author{Geoffroy Lesur\inst{1}, Jonathan Ferreira\inst{1} and Gordon I. Ogilvie\inst{2}   }
   \institute{UJF-Grenoble 1 / CNRS-INSU, Institut de Plan\'etologie et d'Astrophysique de Grenoble (IPAG) UMR 5274, Grenoble, F-38041, France \\
              \email{geoffroy.lesur@obs.ujf-grenoble.fr}
              \and
              Department of Applied Mathematics and Theoretical Physics, University of Cambridge, Centre for Mathematical Sciences,
Wilberforce Road, Cambridge CB3 0WA, UK
                     }

   \date{Received date / Accepted date}

% \abstract{}{}{}{}{} 
% 5 {} token are mandatory
 
  \abstract { Magnetorotational turbulence and magnetically driven disc winds are often considered as separate processes. However, realistic astrophysical discs are expected to be subject to both effects, although possibly at different times and locations. We investigate here the potential link between these two phenomena using a mixed numerical and analytical approach. We show in particular that large-scale MRI modes which dominate strongly magnetised discs (plasma $\beta\sim 10$) naturally produce magnetically driven outflows in the nonlinear regime. We show that these outflows share many similarities with local and global disc wind solutions found in the literature. We also investigate the 3D stability of these outflows and show that they are unstable on dynamical timescales. The implications of these results for the transition between a jet-emitting disc and a standard ``viscous'' disc are discussed.  }


   \maketitle
%
%________________________________________________________________

\section{Introduction}

Accretion discs are found around several kind of astrophysical objects: from young stars (protoplanetary discs) to supermassive black holes in AGNs. The dynamics of these discs is however poorly understood. It is known that, on average, matter moves inward resulting in the accretion of material onto the central object. Dynamically, one can show easily \citep{LP74} that this accretion is controlled by the loss of angular momentum of the falling gas. Hence, if one is to predict the accretion rate and survival time of discs, one needs to understand the angular momentum dynamics of these objects.

For accretion to happen on timescales compatible with observations, angular momentum must be transported rather efficiently. This has led to the $\alpha$ disc model \citep{SS73} in which angular momentum is transported radially outward in the disc by a prescribed ``turbulent viscosity''. The origin of this turbulent viscosity has been debated for several decades and remains even today rather unclear. The magnetorotational instability (MRI: \citealt{BH91a,HGB95}) appears as the most promising way of producing turbulence in discs, though other processes might also be at work in some regions (see \citealt{A11} for a review). However, transporting angular momentum in discs is not the only way to cause accretion. Instead, one can suppose that angular momentum is ``extracted'' from a disc by a large-scale magnetic structure anchored in the disc midplane \citep{BP82,PN83}. This sort of mechanism usually produces magnetically driven outflows, also known as disc winds. 

It is most likely that these two mechanisms (angular momentum transport through turbulence and extraction by winds) actually coexist in astrophysical objects. However, this link is, from a theoretical point of view, poorly understood. The first reason is that stationary disc wind models require a strong magnetisation $\beta\sim 1$ \citep{FP95} which quenches the MRI by magnetic tension effects \citep{BH91a}. At first sight, the conditions of existence for MRI turbulence and disc winds are therefore mutually exclusive. Moreover, a numerical computation including both a large-scale wind and small-scale turbulence in the disc midplane is technically very challenging: resolving turbulence in the bulk of the disc requires a large resolution in the disc midplane and fully 3D simulations, whereas computing the wind itself requires a very large computational domain. Several authors have therefore computed disk winds using a prescription to take into account the effects of disc turbulence (essentially a turbulent viscosity and resistivity), both analytically \citep{CF00} and numerically \citep{ZF07}. On the MRI side, most of our knowledge comes from shearing box simulations \citep{HGB95,SHGB96,LL10}, although several authors have also considered global configurations \citep{H00,FL11,FD11}. These global configurations are however limited to situations without any poloidal magnetic structure (or very weak mean poloidal magnetic fields) which precludes the production of outflows.

More recently the production of outflows from MRI turbulence in shearing boxes has been studied in the limit of weak poloidal fields, $\beta=10^5-10^4$ \citep{SI09}. A magnetocentrifugal mechanism similar to \cite{BP82} has been identified in these simulations, with non-steady outflows starting about two scale heights above the midplane \citep{FLLO12}. However, the outflow mass loss rate was shown to depend strongly on the box vertical size and aspect ratio.   Moreover, the outflow for such a weak mean field is dynamically insignificant for the disc as it extracts a very small amount of angular momentum. In the strong poloidal field regime ($\beta \lesssim1$) which is stable to MRI modes, \cite{O12} studied the production of quasi 1D  steady outflows in shearing boxes. This study demonstrated that some properties of global solutions were recovered in shearing boxes, although the procedure used in this work was not used to look in the parameter regime unstable to the MRI.

The aim of this paper is to study the potential link between the MRI and quasi-steady outflows which are found in global models and simulations. To this end, we study stratified shearing boxes threaded by a strong poloidal field ($\beta\sim 10$) which are not far from the steady solutions of \cite{O12} but are lying in the MRI unstable regime. We first present the shearing box model and the numerical methods we have used in this investigation. We then look at the saturation of 1D MRI modes which naturally produce outflows. These ``MRI-outflows'' are compared to other solutions found in the literature, both local and global. We then study the 3D stability of these outflows and demonstrate that they are unstable on dynamical timescales. The consequences of this instability and its nature are briefly discussed. Finally, we summarise our findings and discuss their potential implications for astrophysical outflows and jets.

\section{Local model}
\subsection{Equations\label{equations}}
MRI-related turbulence and the shearing box approximation have been extensively described in the literature. However, since the physics we are looking for involves outflows and mass losses, we recall here briefly the basic equations for the shearing box model. Interested readers may also consult \cite{HGB95}, \cite{B03} and \cite{RU08} for an extensive discussion of this approximation.

The shearing-box equations are found by considering a cartesian box centred at a fiducial radius $R_0$, rotating with the disc at constant angular velocity $\Omega=\Omega_K(R_0)$ and having dimensions $(L_x, L_y, L_z)$ with $L_i\ll R_0$. We define $r-R_0\rightarrow x$, $R_0\phi\rightarrow y$ similarly to \cite{HGB95}. 
Furthermore, we will assume the disc follows an isothermal equation of state with a constant sound speed $c$. In this approximation, the MHD equations read:
\begin{eqnarray}
\label{eq:mass}\partial_t\rho+\bm{\nabla\cdot }\rho\bm{u}&=&0,\\
\label{eq:motion}\partial_t\rho \bm{u}+\bm{\nabla\cdot } (\rho \bm{u\otimes u})&=&-c^2\bm{\nabla}\rho+(\bm{\nabla\times B})\bm{\times B}\\ 
 \nonumber &&-2\rho\bm{\Omega\times u}-\rho \bm{\nabla}{\psi},\\
\label{eq:induction} \partial_t\bm{B}&=&\bm{\nabla\times}(\bm{u\times B}),
\end{eqnarray}
where we have chosen the units so that $\mu_0=1$ and $\psi$ is a local expansion of the effective gravitational potential around $R_0$:
\begin{equation}
\psi=-q\Omega^2x^2+\frac{1}{2}\Omega^2z^2
\end{equation}
in which we have considered a Keplerian disc having a rotation profile $\Omega_K(r)\propto r^{-q}$ with $q=3/2$.

One should note that the above set of equations admits a solution as a linear shear flow which is a local approximation to the Keplerian profile, $\bm{u}=-q\Omega x\bm{e_y}$. In the following, we will consider perturbations $\bm{v}$ (not necessarily small) to this Keplerian profile so that $\bm{u}=\bm{v}-q\Omega x\bm{e_y}$. 

In the following, we consider a shearing box of size $(L_x,L_y)=(16,16)$, the unit of length being defined by the disc scale height: $H=c/\Omega$. The time unit is $\Omega^{-1}$ and the velocity unit is $c$. We will assume $\rho=1$ in the disc midplane at the hydrostatic equilibrium which sets our unit of mass. As shown below, we only consider the upper half of the disc, so that $z\in[0,z_B]$ where $z_B$ is the altitude of the upper $z$ boundary condition. Unless otherwise stated, we assume $z_B=6$. Because the total vertical magnetic flux is conserved, we introduce the dimensionless magnetization 
\begin{equation}
\mu=\frac{B_z^2}{\Sigma \Omega c},
\end{equation}
 as a control parameter\footnote{Note that this parameter is constant because we assume the total mass in the box to be constant} of our simulations, where $\Sigma=\int dV\, \rho/(L_x L_y)$ is the equivalent surface density of the box.

\subsection{Numerical model}
\subsubsection{Numerical method}
In order to investigate the above system of equations, we use the PLUTO code \citep{MBM07}. PLUTO is a finite volume method using a Godunov scheme to integrate the equations in their conservative form. MHD terms are computed using the constrained transport method of \cite{EH88} which enforces $\bm{\nabla \cdot B}=0$ at machine precision during the evolution of the physical system. We use the Roe method to solve the Riemann problem at cell boundaries. This choice was dictated by the presence of strong numerical instabilities with the HLLD solver when the plasma beta $\beta=2\rho c^2/B^2$ was too small (typically $\beta<1$). Moreover, in order to stabilise the code, we switch to an HLL solver (more diffusive) and a minmod slope limiter whenever the magnetisation is very large (typically $\beta<10^{-4}$) in 3D runs. Only a few cells have such a small $\beta$ and the precise value of the threshold does not significantly change the outcome of the simulations. This however prevents the code from failing when the Alfv\'en speed becomes too large. Similar techniques have been used in numerical studies of supersonic interstellar turbulence \citep{LS09}. All the simulations discussed in this paper are summarised in Tab.~\ref{tab:runs}. 

\subsubsection{Boundary conditions}
In order to reduce the computational costs of the simulations, we compute only the upper half of the disc. The lower half is deduced by symmetry:
$\rho(-z)=\rho(z)$ ; $\bm{v}_H(-z)=\bm{v}_H(z)$ ; $v_z(z)=-v_z(z)$ ; $\bm{B}_H(-z)=-\bm{B}_H(z)$ and $B_z(-z)=B_z(z)$ where $H$ stands for the horizontal $(x,y)$ component of a vector. It should be noted that this symmetry is a natural symmetry of the equations of motion. This implies that if the initial conditions satisfy this symmetry, the resulting solution will verify this symmetry at all times.

The boundary conditions we impose are shear-periodic in the radial direction and periodic in the azimuthal direction. The midplane symmetry described above is imposed at $z=0$. The upper boundary condition ($z=z_B$) is the most delicate part of the setup. Unless otherwise stated, we consider modified outflow boundary conditions where we enforce a strictly vertical poloidal field at the boundary:
\begin{eqnarray}
\partial_z \rho(z_B)=\partial_z \bm{v}_H(z_B)=\partial_z B_y(z_B)&=&0\\
B_x(z_B)&=&0
\end{eqnarray}
Surprisingly, strict outflow boundary conditions (zero gradient for all fields) prevent MHD driven winds to be produced. This point is discussed more extensively in \S\ref{sec:upperbound}. Boundary conditions used for each run are specified in Tab.~\ref{tab:runs}.

Because many of the simulations described here show an MHD-driven wind, a significant amount of mass is lost in our model. In order to mimic the mass inflow due to the material which would be accreted in a realistic disc including curvature effects, we have chosen to add mass in the midplane to compensate for wind losses. This is accomplished by adding mass at $z<z_\mathrm{inj}$ at each numerical time step. This procedure is in a sense a way to artificially introduce curvature effects in the shearing box, but it is clearly unsatisfactory. In particular, conservation of energy and momentum \emph{in the injection region} $z<z_\mathrm{inj}$ are not satisfied, which is clearly an issue if one considers the disc wind problem globally. In all the simulations discussed below all have $z_{\mathrm{inj}}=0.1$, unless otherwise stated. Tests with $z_{\mathrm{inj}}=0.05$ have shown that none of the result we discuss thereafter are significantly affected.

\subsubsection{Modified potential}
The vertical hydrostatic equilibrium described by (\ref{eq:motion}) leads to a gaussian vertical density profile:
\begin{equation}
\label{eq:hydroz_eq}
\rho(z)=\exp\Big[-\frac{z^2}{2H^2}\Big]
\end{equation}
Assuming a constant $B_z$ in the box we get an Alfv\'en speed $V_A=B_z/\sqrt{\rho}\propto\exp(z^2/4H^2)$ which increases very steeply as $z$ increases. Because of the CFL condition, this leads to very small timesteps which dramatically increase the computational time. To reduce the computational costs, several of our simulations were performed using the modified potential:
\begin{equation}
\psi'=-q\Omega^2x^2+\frac{z_0^2}{2}\Omega^2\Big[1-\exp(-z^2/z_0^2)\Big]
\end{equation}
This modified potential is roughly identical to the real potential for $z<z_0$ but is less steep for $z>z_0$ leading to a smoother density profile in the hydrostatic equilibrium and therefore smaller Alfv\'en velocities. It should be noticed that this problem arises only in the hydrostatic equilibrium \emph{without outflow}. As we will show below, when an outflow is produced, the density profile is much smoother and we do not need a modified potential anymore.

We have used the above modified potential with $z_0=4$ in our 1D simulations to initiate the MHD wind in the tall box simulations ($z_B\ge 8$). Once the quasi steady wind was formed, we relaxed back the potential to the original shearing box potential. Comparisons between the solutions obtained with and without the modified potential in the case $z_B=6$ have shown no difference once the potential has been relaxed (runs \textsc{1DRef}---\textsc{1Dz6}).

\section{One-dimensional MRI outflows\label{sec:1Dout}}
In this section, we look at one-dimensional solutions of the equations (\ref{eq:mass})---(\ref{eq:induction}) i.e. ($\rho(z,t)$, $\bm{v}(z,t)$, $\bm{B}(z,t)$). This simplification allows us to isolate the physical mechanisms responsible for the MHD-driven outflows.

\begin{figure*}
   \centering
   \includegraphics[width=0.44\linewidth]{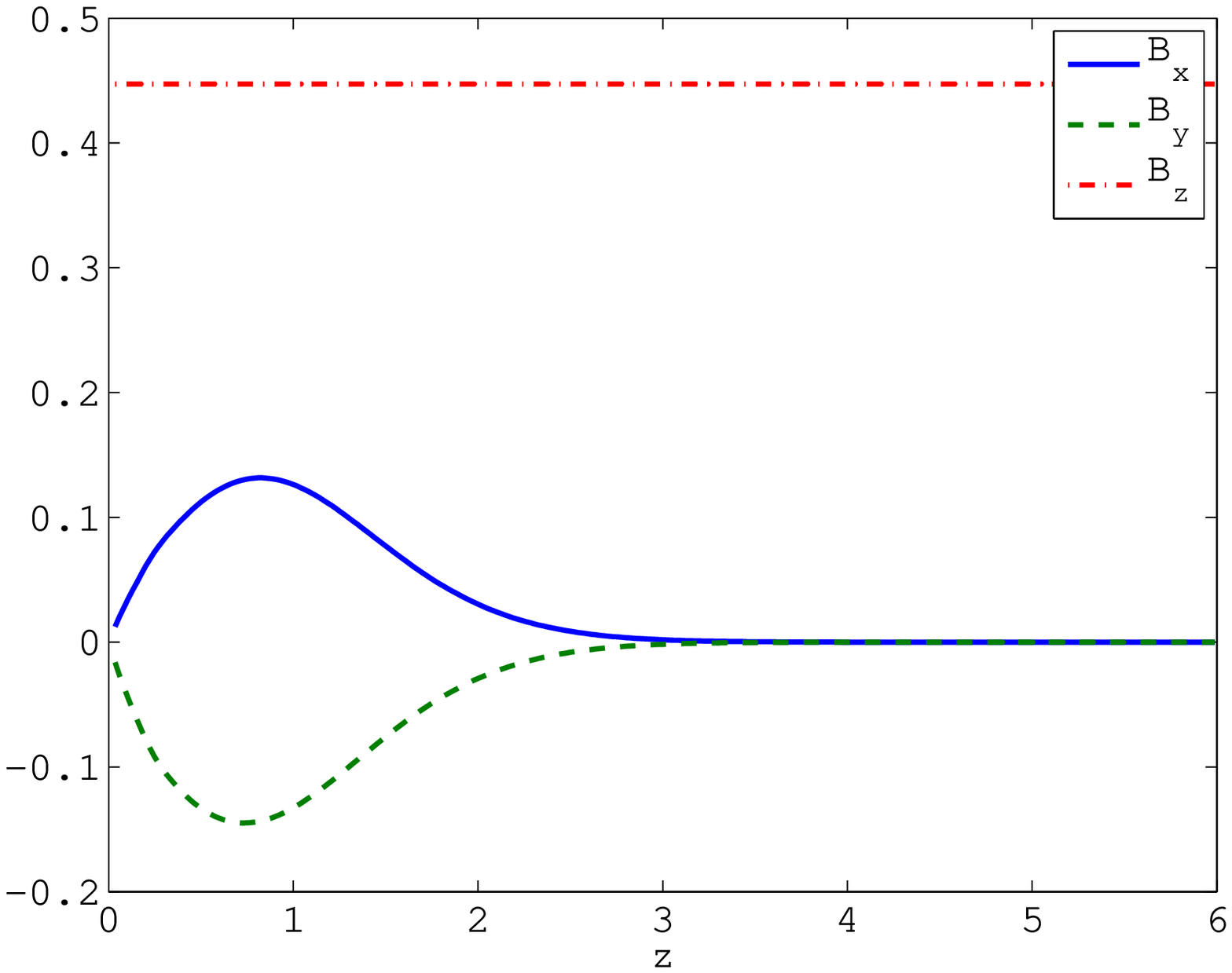}
   \includegraphics[width=0.44\linewidth]{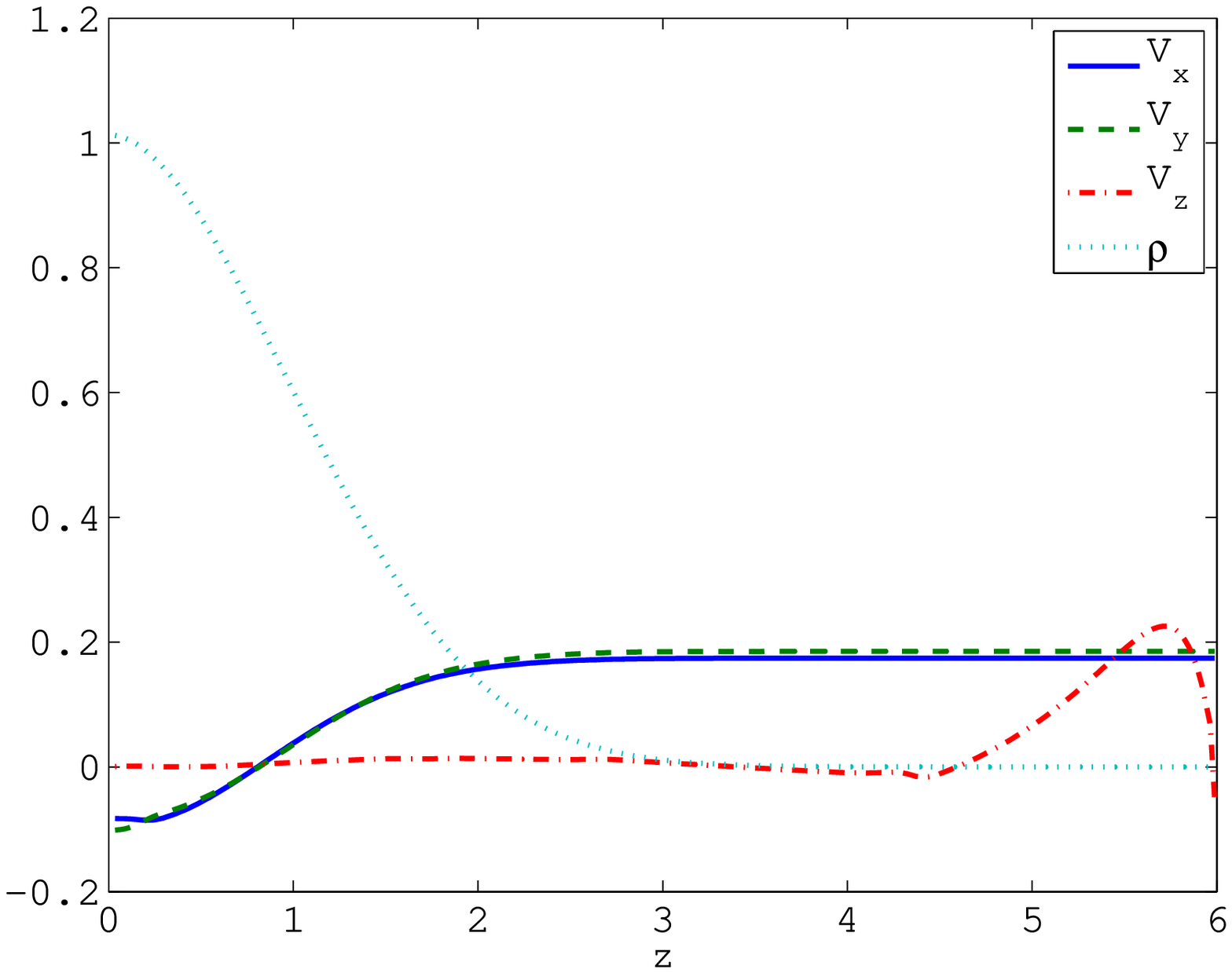}
   \includegraphics[width=0.44\linewidth]{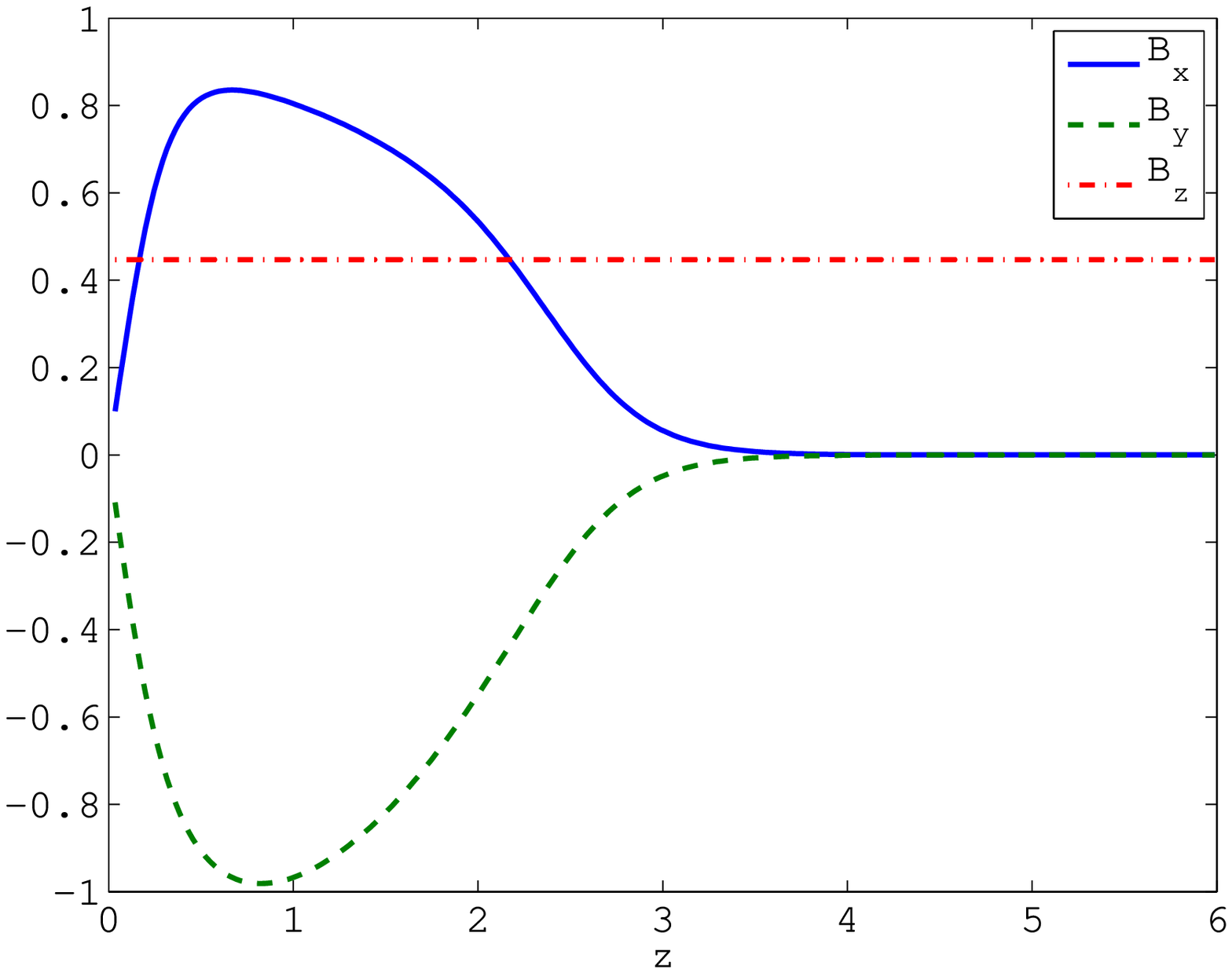}
   \includegraphics[width=0.44\linewidth]{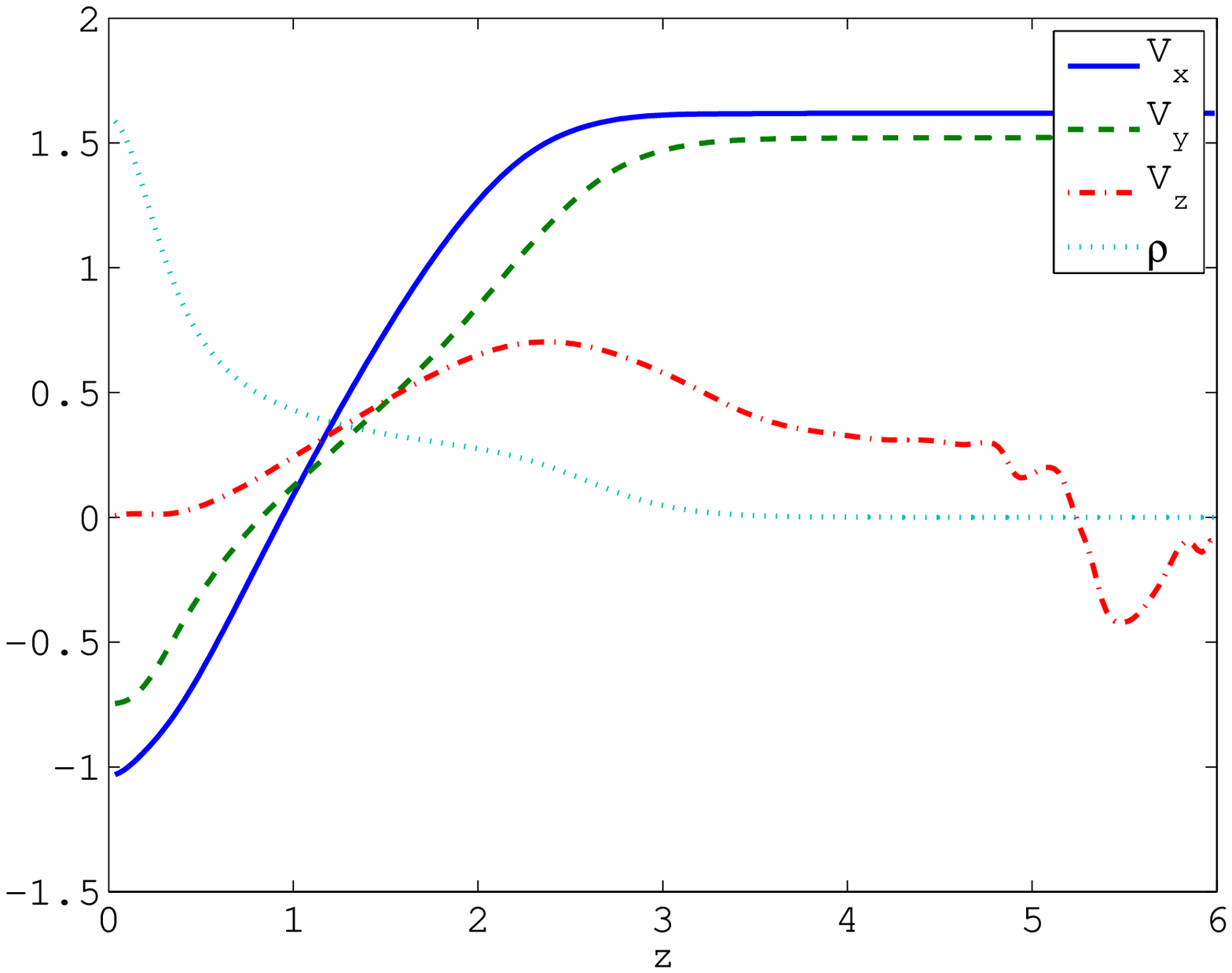}
   \includegraphics[width=0.44\linewidth]{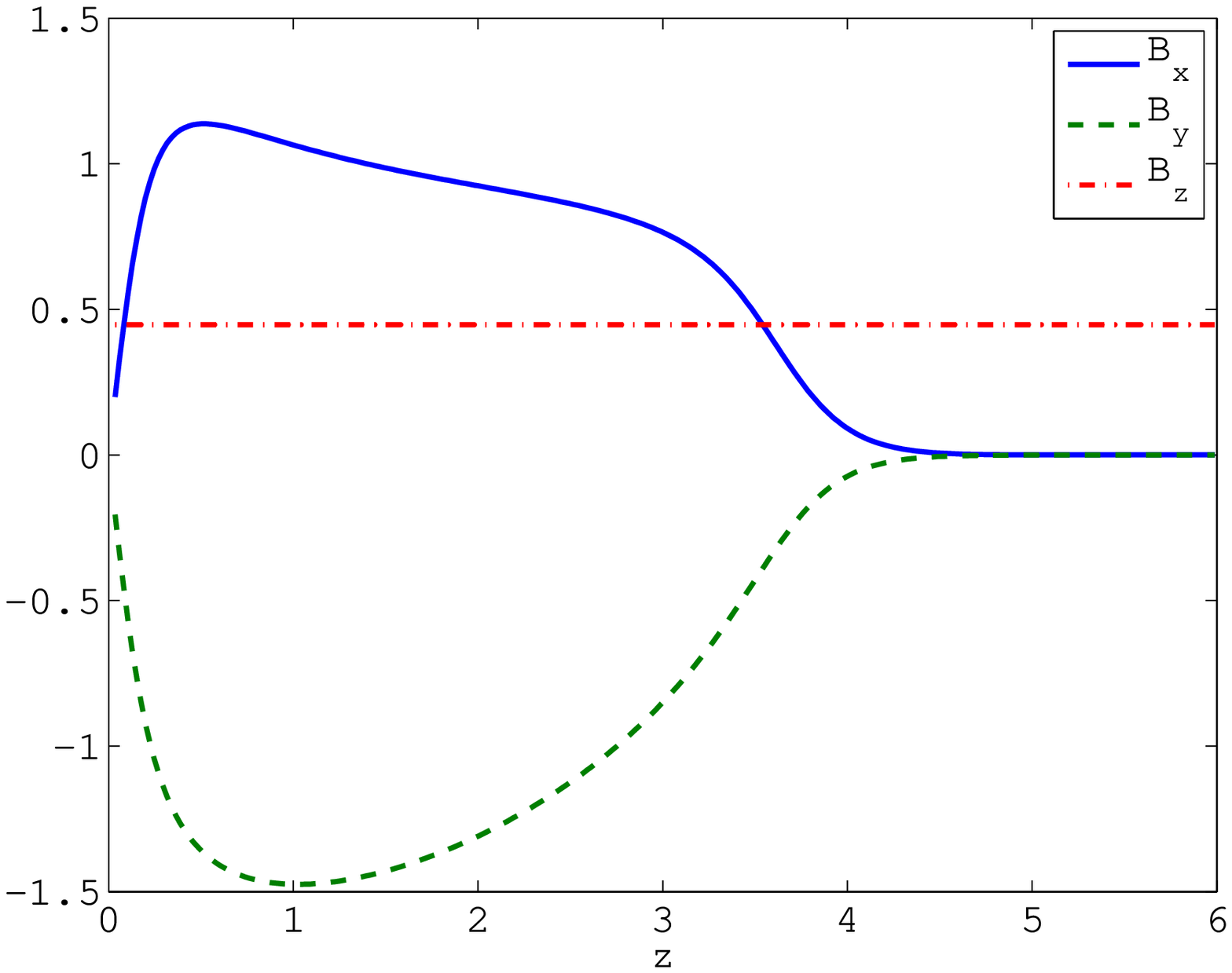}
   \includegraphics[width=0.44\linewidth]{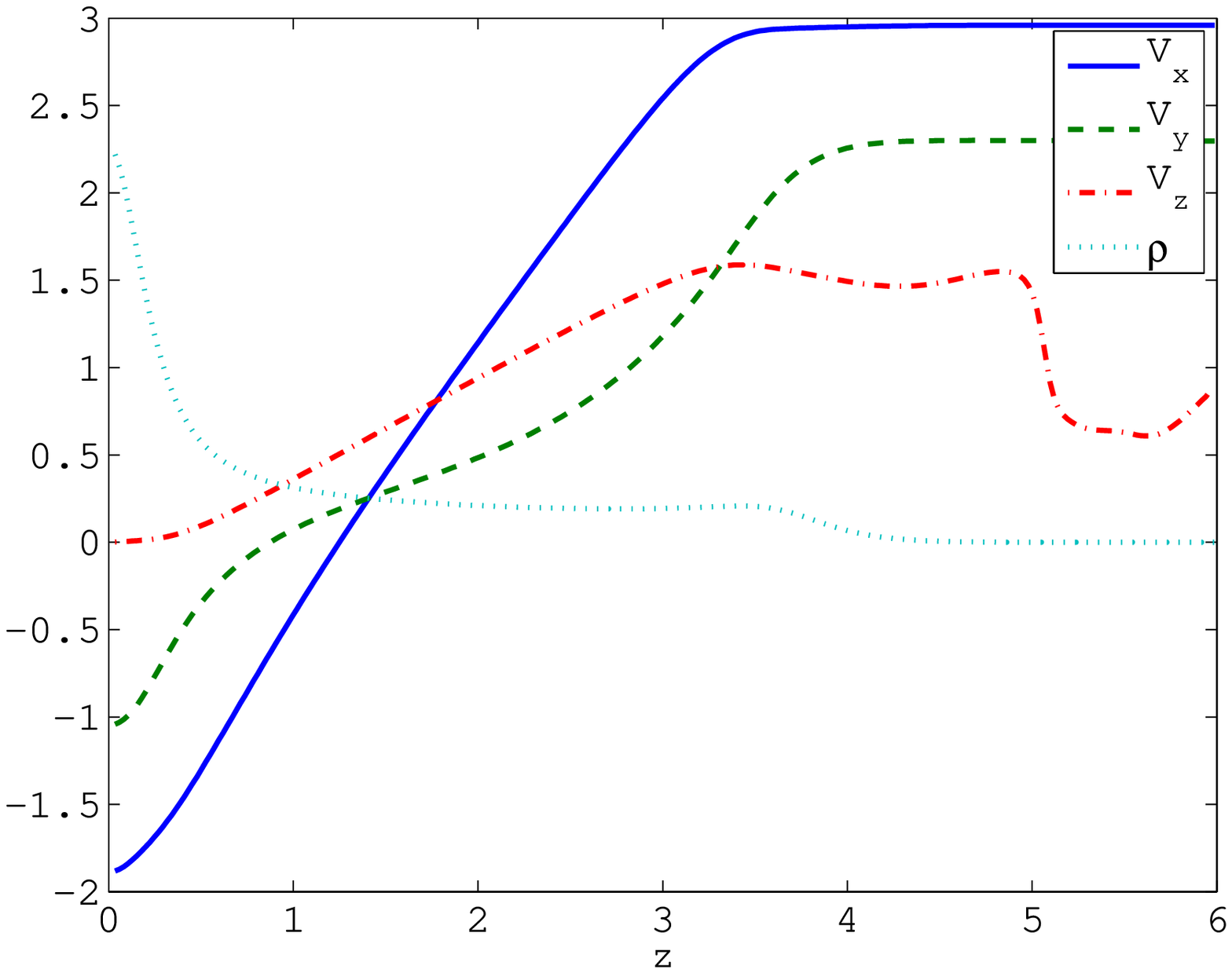}
   \includegraphics[width=0.44\linewidth]{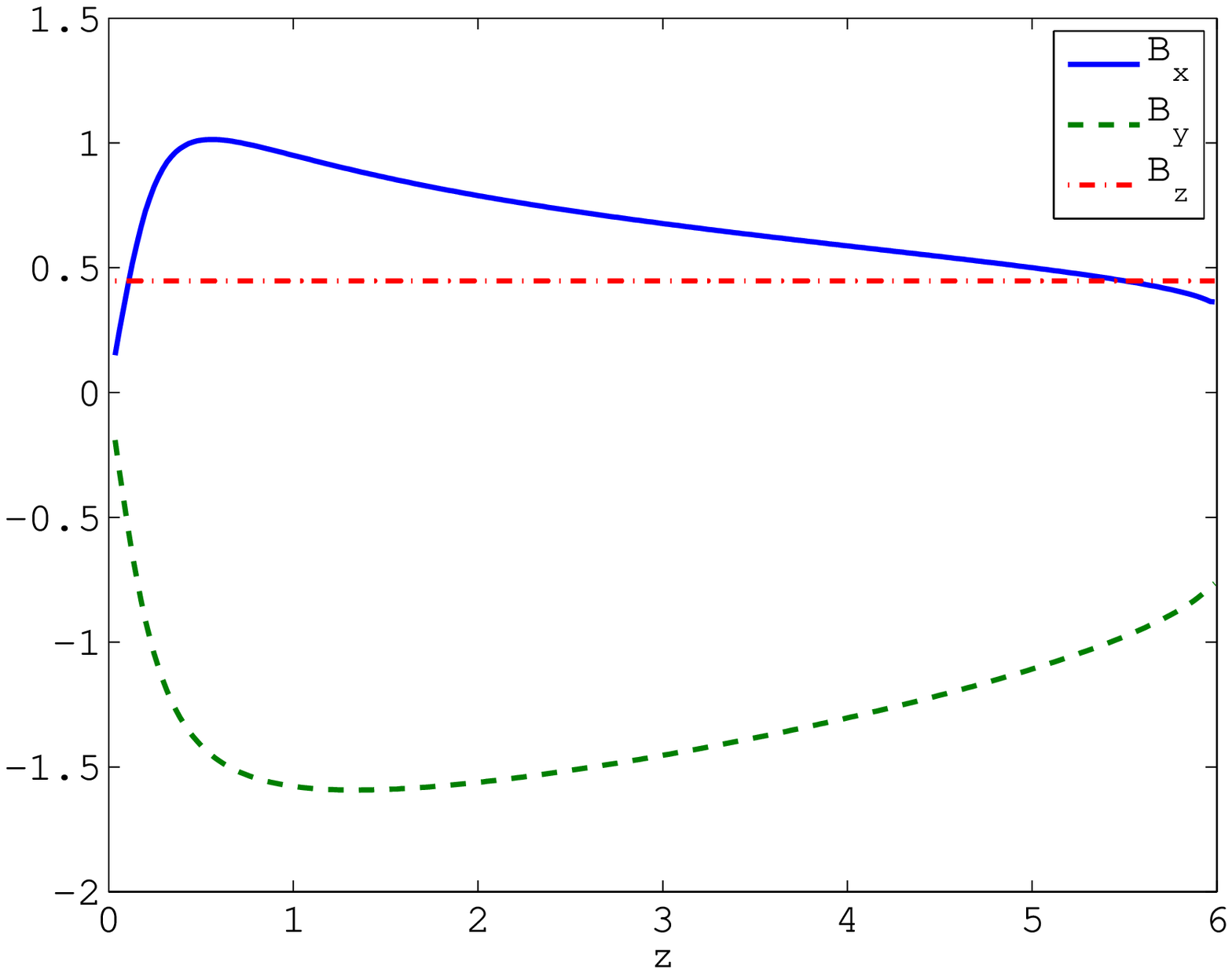}
   \includegraphics[width=0.44\linewidth]{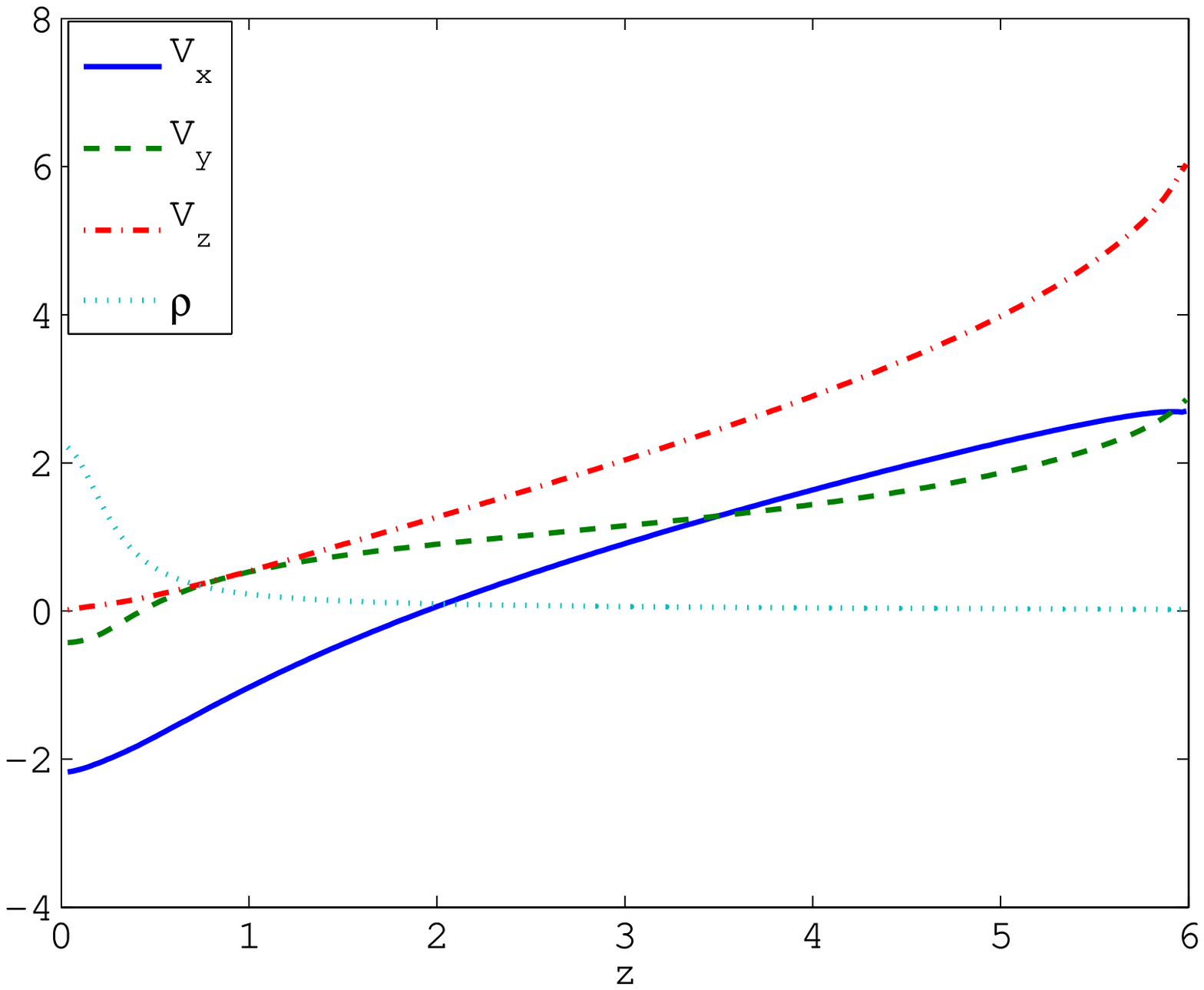}
   
   \caption{Time evolution of the fiducial 1D model \textsc{1DRef}. Left: magnetic field profile, right: Velocity and density profile. From top to bottom: $t=4.0$ ; $7.0$ ; $8.0$ ; $60.0$ }
              \label{Fig:profiles}%
    \end{figure*}

We initialise our box with a disc in vertical hydrostatic equilibrium (\ref{eq:hydroz_eq}). We add a mean vertical field $B_z$ so that $\mu=8\times 10^{-2}$. In order to initialise the growth of MRI modes, we add a small perturbation $B_x=0.02\sin(z)$ to the system. We show the temporal evolution of this run (\textsc{1DRef}) in Fig.~\ref{Fig:profiles}.

\subsection{From MRI modes to steady outflows\label{sec:steady_flow}}
\begin{figure}
   \centering
   \includegraphics[width=0.99\linewidth]{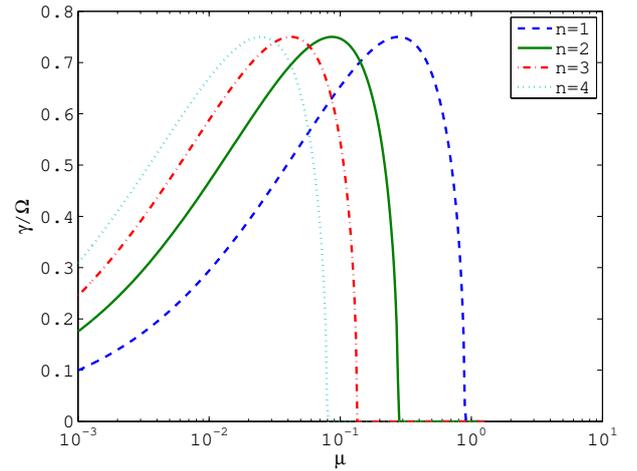}
   
   \caption{Growth rates of the largest stratified MRI eigenmodes as a function of the magnetization parameter $\mu$. These growth rates were deduced from eq.~(18) in \cite{LF10}.}
              \label{fig:growth_rate}%
 \end{figure}

At it can be seen from Fig.~\ref{Fig:profiles}, we first observe the development of a linear MRI mode in the simulation box ($t=4.0$). These linear modes were described extensively by \cite{LF10} for stratified shearing boxes. The mode we observe in our particular setup is the $n=2$ mode. This can be easily checked looking at the shape of the perturbation and comparing to the eigenmodes of \cite{LF10}. Moreover, for $\mu=8\times 10^{-2}$, the $n=2$ mode is the most unstable mode in the system (Fig.~\ref{fig:growth_rate}). Because the magnetic fluctuations are localised at $z\sim H$, the magnetic pressure tends to increase at that location, pushing toward the midplane the bulk of the disc and pushing up the disc atmosphere. At $t=7.0-8.0$, the magnetic pressure is sufficiently large to push the atmosphere, creating a ``bubble'' of material. At this stage, the flow is no longer in a linear regime: the eigenmode is modified by nonlinear (e.g. magnetic pressure) terms. In the end, the system relaxes toward a quasi stationary state ($t=60$). This implies that energy which is in injected by the MRI into the eigenmode is evacuated at the same rate in the outflow.

It should be pointed out that we have totally ignored secondary instabilities in this description. These secondary instabilities are usually $x,y$ dependent modes which grow on the top of MRI eigenmodes \citep{GX94,LLB09,PG09,P10}. It is widely believed that parasitic modes are responsible for the breakup of MRI modes into 3D turbulence, although the exact role they play is still controversial \citep{LL11}. Evidently in our 1D model, parasitic instabilities are inhibited, allowing the primary MRI mode to grow virtually for ever. To allow for the growth of parasitic modes, we have reproduced the same setup in 3D, adding 3D noise at moderate level ($\langle v^2 \rangle\sim 10^{-2}$) to the 1D perturbation described above. This systematically led to the production of an outflow before parasitic instabilities could do anything to the MRI eigenmodes.

Although surprising at first sight, this result can be understood using the phenomenology of parasitic modes. First, it should be noted that the most unstable parasitic modes are usually Kelvin-Helmholtz modes \citep{LLB09} growing on MRI modes. The maximum growth rate of Kelvin-Helmholtz modes can be estimated by the local vertical shear rate. If we consider a primary MRI mode of amplitude $\delta v$ with a characteristic vertical size $\delta l$, then the maximum growth rate of the secondary mode is $\gamma_S\sim \delta v/\delta l$. For the secondary mode to have an impact on the MRI mode, we require $\gamma_S>\gamma$ which implies $\delta v> \delta l \gamma$. Moreover, following \cite{LF10}, we have $\delta b\sim B_{z0} \delta v/\Omega H$ where $\delta b$ is the magnetic perturbation of the MRI mode and $B_{z0}$ is the mean vertical field. Therefore, the parasitic mode can destroy the MRI mode only if
\begin{equation}
\frac{\delta b}{B_{z0}} > \frac{\delta l}{H}\frac{\gamma}{\Omega}
\end{equation}
In our system, the MRI mode characteristic length $\delta l$, growth rate $\gamma$ and the mean field amplitude are all of the order of 1 (in code units, see \S\ref{equations}). This implies that parasitic modes will appear only when $\delta b> 1$. However, as it can be seen in fig.~\ref{Fig:profiles}, the outflow starts when $\delta b\lesssim 1$, i.e. \emph{before} parasitic modes could act on the MRI mode. This result is due to the relatively large magnetisation used in these simulations ($\mu\lesssim 1$) compared to traditional MRI setups ($\mu \lesssim 10^{-3}$) . This large magnetisation implies the production of a large scale MRI mode whose growth rate is of the order of the orbital time scale. This makes the outflow more favourable compared to secondary instabilities. On the contrary, when the magnetisation is small, dominant MRI modes are found at smaller scale (large $n$, see fig.~\ref{fig:growth_rate}). In this case, parasitic modes are favoured and a turbulent flow is obtained.

We should emphasize that the presence of an outflow does not mean that parasitic instabilities are totally absent from this picture. As   we will show later (\S\ref{3Dsolutions}), solutions exhibiting an outflow are themself subject to parasitic instabilities. However, these instabilities have nothing in common with the traditional parasitic instabilities of MRI eigenmodes. 

We have seen above that the evolution of a large-scale MRI mode in a strongly magnetised shearing box leads naturally to the production of a magnetically-driven outflow. This steady outflow is essentially one-dimensional and can be described by ${\bm{v}(z), \rho(z), \bm{B}(z)}$. In the following we will concentrate on the structure of this outflow: phenomenology, critical points, boundary conditions and conserved quantities.

\subsection{Outflow phenomenology\label{sec:pheno}}

As we have shown above, the outflow is primarily produced by the magnetic pressure gradient. The magnetic pressure being maximum at $z\sim 1.5$, it pushes up the outflow at $z>2.0$ but it also compresses the bulk of the disc. An alternative view of this effect can be obtained looking at currents. The outflow is in this case due to horizontal currents which are reversed at $z\sim1.5$. We typically have $J_x>0$ and $J_y>0$ in the bulk of the disc whereas $J_x<0$ and $J_y<0$ in the atmosphere $z>2$. It is important to note that the outflow acceleration can occur only if these currents are non-zero and \emph{change sign}. This remark justifies the absence of any outflow with ``zero gradient'' boundary conditions (see \S\ref{sec:upperbound}).

 \begin{figure}
   \centering
   \includegraphics[width=1.1\linewidth]{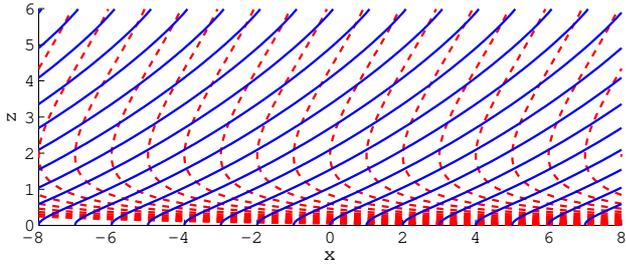}
   
   \caption{Streamlines (red dashed lines) and field lines (blue plain lines) of our steady solution obtained at $t=95$. }
              \label{fig:streamlines}%
 \end{figure}

It is of interest to put these outflow solutions in the context of the \cite{BP82} disc wind paradigm. In this model, the outflow is initiated by a magnetocentrifugal effect: the poloidal magnetic field lines are considered as rigid wires anchored in the bulk of the disc and fluid particles are allowed to drift along these wires. If the field lines are sufficiently inclined (typically more than $30^\circ$ with respect to the vertical axis), then the particles are azimuthally accelerated by the anchored field lines. This leads to a centrifugal effect which ejects fluid particles along field lines. In this picture, the ejection is driven by an exchange of angular momentum: angular momentum is taken from the disc by the field line and it is then released in the ejected material.

In order to compare this mechanism to outflow solutions driven by MRI modes, we show in Fig.~\ref{fig:streamlines} the poloidal streamlines and field lines of such a solution. We first note that poloidal streamlines and fieldlines are not aligned. This property is allowed by the shearing box boundary conditions, but it physically means that magnetic flux is ``accreted'' toward the centre of the disc\footnote{In a shearing box, the flux coming in the box through the $x=+L_x/2$ boundary condition is equal to the flux leaving the box through the $x=-L_x/2$ boundary condition, making the magnetic field configuration overall stationary. Such a solution is however very specific to the shearing box and does not represent a steady situation in an accretion disc}. In a more realistic model including curvature, such a state could not be sustained for a long time because magnetic flux would get accumulated at the disc centre, thereby modifying the disc properties (especially its rotational profile). This is the principle motivation for the presence of a strong ``magnetic diffusivity'' in disc wind models, either assuming the presence of small scale turbulence \citep{FP93}, ambipolar diffusion \citep{WK93} or Hall and Ohm diffusion \citep{KSW10}.

Despite this difference, we recover most of the phenomenological properties of the \cite{BP82} paradigm: field lines are inclined and drive an outflow which is inclined towards the same direction. This indicates that angular momentum is exchanged between the field and the flow. As we will see below (\S\ref{sec:angmom}), angular momentum is effectively taken away from the disc by the field and then released to the ejected material. This effect is actually inevitable since the current configuration described previously inevitably leads to a positive magnetic torque $\propto J_z B_x-J_xB_z$ in the outflow (and a negative one in the disc midplane). Because angular momentum is taken from the disc by the field, a strong radial flow is produced which explains the streamlines' orientation for $z<1$. Finally, we find an inclination angle of $\sim40^\circ$ for the poloidal magnetic field lines and $\sim 25^\circ$ for the poloidal velocity field. This last value is very close to the critical value of \cite{BP82}.

\subsection{Critical points}

In principle, it is possible to look systematically for a steady 1D solution of equations (\ref{eq:mass})---(\ref{eq:induction}). This is done writing the equations of motion in the form $\mathbf{M\cdot X}=\mathbf{Y}$, where 
\begin{equation}
\label{eq:nonlinear1Da}
\mathbf{M} = \left( \begin{array}{cccccc}
v_z & 0 & 0 & \rho & 0 & 0 \\
0 & \rho v_z & 0 & 0 & -B_z & 0 \\
0 & 0 & \rho v_z & 0 & 0 & -B_z \\
c^2 & 0 & 0 & \rho v_z & B_x & B_y \\
0 &-B_z & 0 & B_x & v_z & 0 \\
0 & 0 & -B_z & B_y & 0 & v_z
\end{array}\right)
\end{equation}
\begin{eqnarray}
\label{eq:nonlinear1Db}
\mathbf{X}=\partial_z\left(\begin{array}{c}
\rho \\
v_x\\
v_y\\
v_z\\
B_x\\
B_y
\end{array}
\right) &;& \mathbf{Y}=\left(\begin{array}{c}
0 \\
2\Omega\rho v_y\\
-(2-q)\Omega\rho v_x\\
-\rho\Omega^2z\\
0\\
-q\Omega B_x
\end{array}
\right)
\end{eqnarray}
where $\mathbf{M}$ and $\mathbf{Y}$ are a matrix and a vector which do not contain any spatial derivative. In order to solve this nonlinear system, one then inverts the above system of equation to get a set of ordinary differential equations $\mathbf{X}=\mathbf{M}^{-1}\cdot \mathbf{Y}$. However, when a critical point is reached, $\mathbf{M}$ is singular and the system cannot be inverted anymore. In this case, the physical system needs to satisfy an extra condition in order to get through the critical point.

In the shearing box, we find three types of singular points: 
\begin{itemize}
\item two slow magnetosonic points
\begin{equation}
v_z^2=\frac{1}{2}\Big[V_A^2+c^2-\sqrt{(V_A^2+c^2)^2-4c^2V_{Az}^2}\Big],
\end{equation}
\item two Alfv\'en points 
\begin{equation}
v_z^2=V_{Az}^2,
\end{equation}
 \item and two fast magnetosonic points 
 \begin{equation}
v_z^2=\frac{1}{2}\Big[V_A^2+c^2+\sqrt{(V_A^2+c^2)^2-4c^2V_{Az}^2}\Big],
 \end{equation}
 \end{itemize}
where $V_A=\sqrt{\bm{B}^2/\rho}$ and $V_{Az}=B_z/\sqrt{\rho}$. In the following, thanks to symmetries, we will only consider solutions with $v_z>0$ so that only one critical point of each kind will be present.

\begin{figure}
   \centering
   \includegraphics[width=1.0\linewidth]{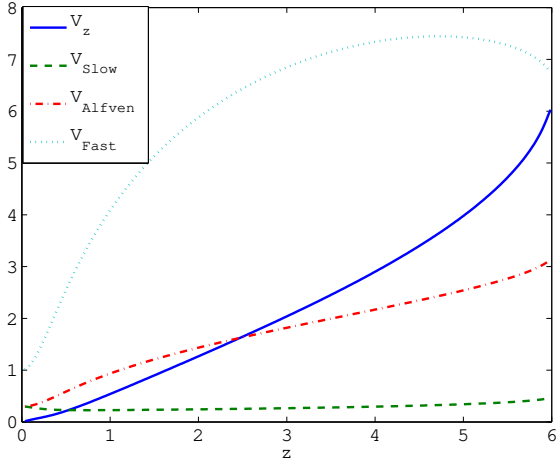}
   
   \caption{Vertical velocity and MHD wave speeds for the fiducial outflow solution at $t=60$.  }
              \label{fig:critpoints}%
 \end{figure}

We present the MHD wave speeds and flow speeds in Fig.~\ref{fig:critpoints}. We find that the slow point is located around $z=0.52$ and the Alfv\'en point is found at $z=2.47$. The flow does not cross the fast point, however we find that the fast speed and the flow vertical speed tend to converge more rapidly close to the upper boundary. Note that a very similar behaviour was observed in global self-similar solution close to the fast surface (\citealt{CF00}, Fig.~1). This result indicates that the flow is still causally connected to the disc and therefore the boundary condition we impose at the top of the box still has an impact on the flow structure itself. This point will be discussed in the next section.

\subsection{Influence of the upper boundary condition}
\label{sec:upperbound}
We have seen above that the outflow is still physically connected to the disc since it is not super-fast. Moreover, it looks as if the fast magnetosonic point is located close to the imposed upper boundary condition. One might wonder what is the exact role played by this boundary condition. In order to investigate this issue, we have performed two kinds of test, either modifying the altitude $z_B$ of the vertical boundary condition or modifying the nature of the upper boundary condition we apply.

First, varying the altitude of the vertical boundary condition $z_B$ led to the results presented in Tab.~\ref{tab:bc}. One finds that modifying the altitude of the boundary condition strongly modifies the outflow properties. In particular, we find that increasing the altitude of the boundary condition leads to a decrease in the flow ejection rate. This evolution is accompanied by a slow point moving closer to the midplane and an Alfv\'en point moving higher up in the atmosphere. This result clearly demonstrates that because the flow has not crossed the fast magnetosonic point, the solution we obtain is still constrained by the boundary condition we impose at $z_B$. In all the solutions described in Tab.~\ref{tab:bc}, we have observed a ``convergence'' of the fast magnetosonic speed and the flow speed when approaching $z_B$, similarly to what we find in Fig.~\ref{fig:critpoints}. This surprising result tends to indicate that our boundary conditions somehow force the fast point to be close to the upper boundary. 

The fact that the slow point moves closer to the disc midplane when the mass loss rate decreases might look dubious to readers familiar with the phenomenology usually associated with the slow point. In particular, it is often said that the slow point ``sets'' the wind escape speed through the relation $\rho v_z=\rho_0(z_s) c$ where $\rho_0$ is taken as the hydrostatic density profile \citep{S96}. This argument predicts that smaller mass loss rate should be associated to slow points located higher up in the atmosphere, which is exactly the opposite of what we find. This simple argument is however not valid in the present case since the density profile deviates \emph{significantly} from the hydrostatic equilibrium. In particular, configurations \textsc{1Dz16} and \textsc{1Dz20} exhibit strongly compressed discs due to a large magnetic pressure gradient in the atmosphere which most probably prevents significant ejection from happening.

\begin{table}
\centering
\begin{tabular}{|l|l|l|l|l|c|c|}
\hline
Run & $z_B$ & $\rho v_z$ & $z_S$ & $z_A$ & $\theta_B(z_A)\,({\,}^\circ)$ & $\theta_V(z_A)\,({\,}^\circ)$ \\
\hline 
\textsc{1Dz4} & 4         &  0.147   &   0.583               &    2.03      &     56.8  &  8.0     \\
\textsc{1Dz6} & 6         &  0.123   &   0.530               &    2.47      &     58.6   &  16.4    \\
\textsc{1Dz8} & 8         &  0.105   &   0.509               &    2.94      &     58.8   &  22.1    \\
\textsc{1Dz12} & 12       &  0.086   &   0.483               &    3.83     &     58.7   &   28.0    \\
\textsc{1Dz16} & 16       &  0.072	   &   0.475               &    4.77    & 57.9    & 31.4           \\
\textsc{1Dz20} & 20       &  0.063	   &   0.470               &    5.86   &   56.5   &  32.8        \\
\hline
\end{tabular}

\caption{\label{tab:bc}Evolution of the mass loss rate $\rho v_z$, slow point $z_S$ and Alfv\'en point $z_A$ as a function of the altitude of the boundary condition $z_B$. We also show the inclination angle with respect to $z$ of the poloidal field lines and stream lines at the Alfv\'en point ($\theta_B(z_A)$, $\theta_V(z_A)$)}
\end{table}

We have also tried to modify the nature of the upper boundary conditions. First, instead of imposing $B_x(z_B)=0$, we have imposed a fixed angle to the poloidal field, i.e. $B_x(z_B)=\tan(\theta)B_z$ with $\theta=30^\circ\,;\,45^\circ$ as in \cite{O12}. Surprisingly this did not modify significantly the outflow solution we obtained: the field values are modified by less than $5\%$. This can be explained by the fact that the outflow is super-Alfv\'enic when it reaches the top boundary. As we will see below (\S\ref{sec:mag_dependency}), sub-Alfv\'enic outflows are effectively very sensitive to the field configuration at the boundary, but super-Alfv\'enic outflows are not. We conclude from this that the inclination angle of the poloidal field line is set by the Alfv\'en point crossing condition. This result is corroborated by the constant inclination angle at the Alfv\'en surface found when one changes the box vertical size (Tab.~\ref{tab:bc})

 We have finally tried to impose a zero gradient condition on $B_x$ and $B_y$ (classically called ``outflow'' boundary condition). This results in the suppression of the outflow solution. We observe instead a constant increase of the magnetic pressure in the atmosphere which results in a strong compression of the disc material in the midplane until the disc occupies one numerical grid cell. This result is similar to the low $\beta$ simulations of \cite{HGB95} with mean vertical flux. This was to be expected since the outflow is driven by horizontal currents. Imposing a zero gradient condition means that no horizontal current is allowed at the boundary, blocking any potential outflow by cancelling the vertical component of the Lorentz force and the change of sign of the magnetic torque \citep{FP93b}.
 
\subsection{Conserved quantities}
In the following, we assume all the quantities $\bm{v},\quad\bm{B},\quad\rho$ depend only on $z$, as found in the steady ejection above. With this hypothesis, we reconstruct the conserved quantities used in global disc wind models \citep{BP82,PP92,CF00}.
\subsubsection{Frozen field condition}
We first note that under the above conditions, $B_z$ and $\rho v_z$ are constant in the box thanks to flux and mass conservation. We therefore introduce a proportionality constant between these quantities:
\begin{equation}
\rho v_z=\alpha B_z
\end{equation}
Thanks to the $x$ induction equation, we also have $v_x B_z-v_zB_x=\mathcal{E}_y=\mathrm{const}$, which can be simply written as:
\begin{equation}
v_x^*=v_x -v_z\frac{B_x}{B_z}
\end{equation}
where $v_x^*$ is a constant. Physically, $v_x^*$ is the advection speed of the poloidal magnetic field. In global models, this quantity is implicitly set to $0$ in order to avoid flux accumulation at the disc inner edge and potential singularities at $R=0$ \citep{C56,M61,PP92}. This is not required here as we are using a shearing-box model. However, \emph{one should keep in mind that our model implicitly allows radial advection of magnetic field lines}.
The existence of this conserved quantity allows us to define a relation between the poloidal field and the poloidal velocity, namely:
\begin{equation}
\bm{v_p}=\frac{\alpha}{\rho}\bm{B_p}-v_x^*\bm{e_x},
\end{equation}
where $\bm{v_p}$ and $\bm{B_p}$ are the poloidal $(x,z)$ components of the velocity and magnetic fields. This equation shows a major difference between the global approach and the local approach. In the global approach, $v_x^*=0$ (no advection of magnetic field lines), which implies that poloidal streamlines and magnetic field lines are aligned. This is not the case in our solutions, for which $v_x^*=v_x(z=0)<0$.

\subsubsection{Magnetic surface rotation}
A relation, known as the magnetic surface rotation speed can be deduced from the $y$ component of the induction equation. Since $B_y$ does not depend on $x$, we get after some algebra
\begin{equation}
\bm{B_p\cdot\nabla} \Big[ \frac{\alpha}{\rho} B_y-v_y+q\Omega x\Big]=0
\end{equation}
Note that the above relation is only valid along a poloidal magnetic field line. In contrast to \cite{PP92}, no equivalent relation is found along poloidal streamlines. This allows us to define a new conserved quantity
\begin{equation}
v_y^*=v_y-q\Omega x - \frac{\alpha}{\rho} B_y
\end{equation}
which is conserved along a magnetic field line. It should be pointed out at this stage that $\mathcal{E}_x=u_yB_z-u_zB_y=B_z v_y^*$ where $\mathcal{E}_x$ is the $x$ component of the electromotive force. Therefore, in the frame translating at $v_y^*$, the $x$ component of the electromotive force is zero, which justifies the naming of this invariant.

\subsubsection{Angular momentum conservation\label{sec:angmom}}
The angular momentum conservation equivalent is derived from the $y$ component of the equation of motion (\ref{eq:motion}). It should first be noted that this equation can be written:
\begin{equation}
\rho\bm{ v_p\cdot \nabla} \mathcal{L}-\bm{B_p\cdot\nabla}B_y=0
\end{equation}
where $\mathcal{L}=v_y+(2-q)\Omega x$ is the local equivalent of the specific angular momentum.
Since $B_y$ does not depend on $x$, we can rewrite the above equation using the flux freezing condition:
\begin{equation}
\label{eq:angmom}
\rho \bm{v_p \cdot \nabla} \Bigg[\mathcal{L}-\frac{B_y}{\alpha}\Bigg]=0.
\end{equation}
which indicates that $f=\mathcal{L}-\frac{B_y}{\alpha}$ is conserved along streamlines.  

\begin{figure}
   \centering
   \includegraphics[width=1.0\linewidth]{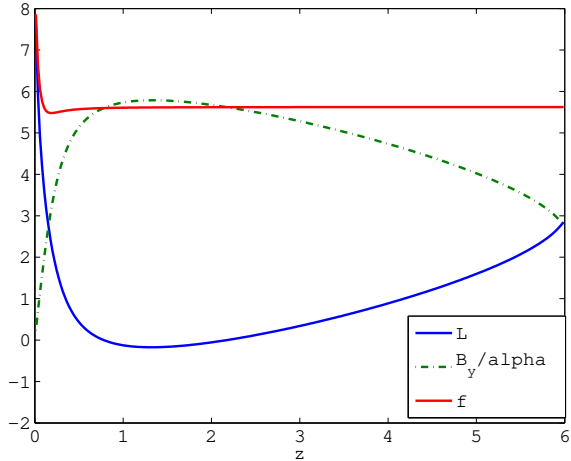}
   
   \caption{Angular momentum along one streamline computed according to (\ref{eq:angmom}). The angular momentum is initially stored in the torooidal field  before being transferred to the flow. }
              \label{fig:angmom}%
              
 \end{figure}

The angular momentum along one streamline is shown in Fig.~\ref{fig:angmom}. This allows us to check that the angular momentum is effectively conserved in our simulations. Moreover, we find that most of the angular momentum is initially extracted from the disc by the toroidal field $B_y$. Higher above the disc ($z\sim 2H$), the angular momentum is exchanged between the toroidal field and the accelerated gas. This demonstrates that the magnetocentrifugal acceleration effect, present in the \cite{BP82} phenomenology, is also found in our outflow solutions.

\subsubsection{Bernoulli invariant}
The Bernoulli invariant is derived from (\ref{eq:motion}) after a scalar multiplication by $\bm{u}$. One gets:
\begin{equation}
\rho\bm{u_p\cdot }\bm{\nabla} \Big[\frac{u^2}{2}+H+\psi\Big]=\bm{J\times B\cdot u}
\end{equation}
where $H$ is the flow enthalpy. The right-hand side of this equation corresponds to the work done by magnetic forces. This term can be calculated as a function of the previously defined invariants which, after some algebra leads to:
\begin{equation}
\label{eq:Bernoulli}\rho\bm{u_p\cdot }\bm{\nabla} \Bigg[\frac{u^2}{2}+H+\psi-\frac{B_y v_y^*-B_x v_x^*}{\alpha}\Bigg]=-q\Omega \rho B_y v_x^*.
\end{equation}
Because $\bm{v_p}$ and $\bm{B_p}$ are not strictly speaking collinear, Bernoulli's equation is left with a non conservative term. However, as we will see later, this term is important only in the bulk of the disc, so that the flux defined above will be \emph{approximately} invariant along a streamline. Comparing this invariant to the one defined in global geometry, we note the presence of a new term in our case, $B_x v_x^*/\alpha$. As shown before, this term describes the energy associated with the field lines being advected. We will see later the role it plays in the ejection process.

\begin{figure}
   \centering
   \includegraphics[width=0.90\linewidth]{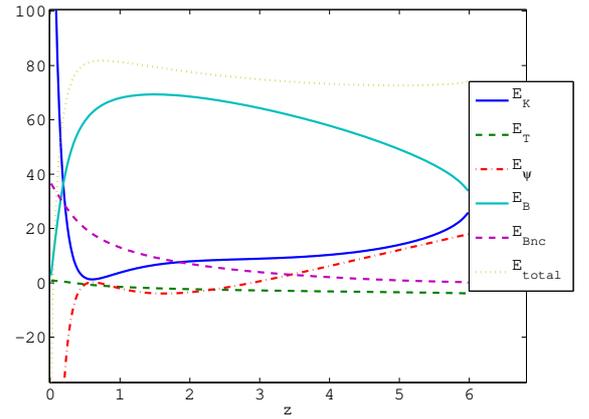}
   
   \caption{Bernoulli invariant computed according to (\ref{eq:Bernoulli}). $E_K$: Kinetic energy, $E_T$: thermal energy (enthalpy), $E_\psi$: Potential energy, $E_B$: Conservative magnetic energy, $E_{Bnc}$: non-conservative magnetic energy.}
              \label{fig:Bernoulli}%
              
 \end{figure}

We show the different terms involved in the Bernoulli invariant (\ref{eq:Bernoulli}) in Fig.~\ref{fig:Bernoulli}. One should note that the magnetic contribution is separated in two parts: (i) the conservative part $(B_y v_y^*-B_x v_x^*)/\alpha$ and (ii) the non-conservative part $\int dl\,q\Omega \rho B_y v_x^*$ where the integral is computed along a streamline.

We first find that the invariant is approximately conserved for $z>0.2$. Initially (up to $z\sim 1.5$) the energy is stored in the conservative component of the magnetic field. The non-conservative part is constantly decreasing, demonstrating that this term is helping the outflow. Higher in the outflow, the magnetic energy is converted into kinetic and potential energy. Finally, we find that the thermal energy plays essentially no role in the ejection energetics. 

The fact that potential energy increases along a streamline might look at first surprising since in a typical \cite{BP82} situation, one would expect the potential energy (gravitational+centrifugal) to \emph{decrease} along a field line. This is not the case here because the inclination angle of the outflowing streamlines is slightly less than $30^\circ$ (see \S\ref{sec:pheno}).

\subsection{Magnetisation dependency}
\label{sec:mag_dependency}

\begin{figure}
   \centering
   \includegraphics[width=1.0\linewidth]{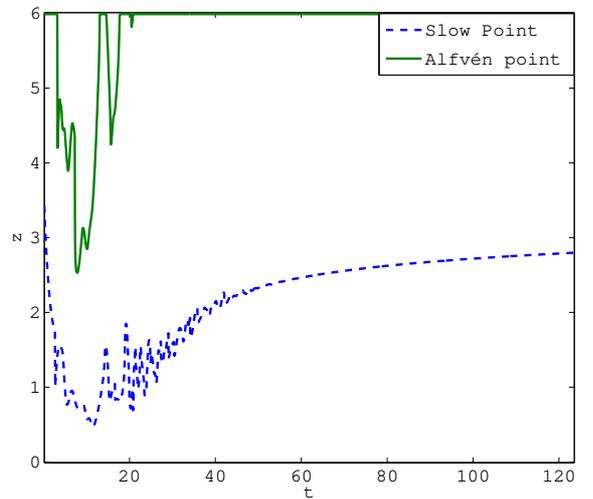}
   
   \caption{Evolution of the critical points location as a function of time in a case without mass injection. The flow becomes sub-Alfv\'enic at $t\simeq 20$ (see text).}
              \label{fig:critpoints_nomass}%
              
 \end{figure}
 
 \begin{figure}
   \centering
   \includegraphics[width=1.0\linewidth]{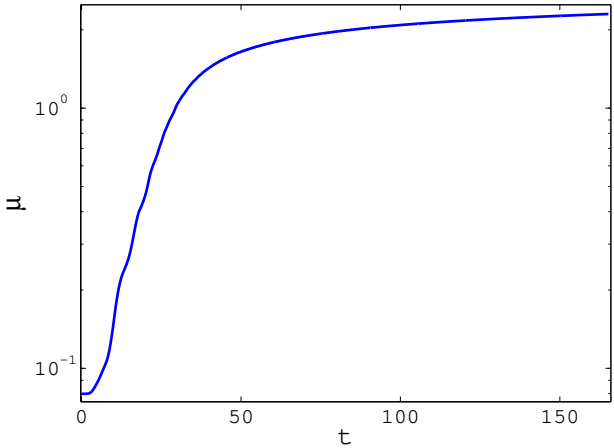}
   
   \caption{Flow magnetisation $\mu$ as a function of time in a case without mass injection. }
              \label{fig:mut_nomass}%
              
 \end{figure}
 
 \begin{figure}
   \centering
   \includegraphics[width=1.0\linewidth]{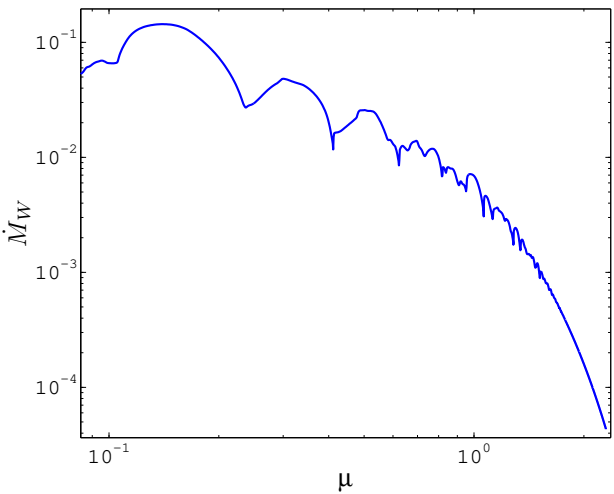}
   
   \caption{Mass loss rate due to the wind as a function of the magnetisation $\mu$. Deduced from a case without mass injection}
              \label{fig:Mw_nomass}%
              
 \end{figure}

In the previous discussion, we kept a constant equivalent surface density $\Sigma$ by artificially injecting mass in the disc midplane. This approximation, although partly motivated by the global disc structure, should be tested more extensively. To do so, we have performed simulations  \emph{without} mass injection. By definition these simulations cannot achieve a steady state. They are however representative of an extreme case in which no mass is coming from the outer disc.

It should first be pointed out that the outcome of these simulations depends strongly on the nature of the upper boundary condition. This is because as the disc mass is lost, the Alfv\'en point moves higher up in the atmosphere (see Fig~\ref{fig:critpoints_nomass}). At some point, the Alfv\'en point crosses the upper boundary and the ejection becomes sub-Alfv\'enic. When this happens in this simplified setting (no radial dependence), the poloidal field inclination is not set anymore by the Alfv\'en point crossing condition (\citealt{O12}, see also \S \ref{sec:upperbound}). Instead, it should be set manually at the upper boundary. If we set $B_x(z_B)=0$, the ejection stops as soon as the Alfv\'en point exits the simulation domain. This is expected since no ejection should be happening with strictly vertical poloidal field lines. On the contrary, if we impose a fixed inclination angle of $45^\circ$ at the upper boundary condition, as in \cite{O12}, the outflow is kept. In the following we will consider the latter case.

Because the disc is losing mass, its magnetisation $\mu$ is increasing as a function of time (see Fig.~\ref{fig:Mw_nomass}). We note however, that the system appears to be approaching a steady state with $\mu\simeq 2.3$. Looking at the correlation between the mass loss rate due to the wind $\dot M_W=(\rho v_z)_{z=z_B}$ and the magnetisation $\mu$ (Fig~\ref{fig:critpoints_nomass}), we find that the mass loss rate decreases steeply for $\mu>1$, which explains the quasi-steady state we observe. We note that this last result is very similar\footnote{Note that our definition of $\dot M_W$ differs from \cite{O12} by a factor $\sqrt{2\pi}$. Moreover, our $\dot M_W$ is normalized by the initial density in the midplane, whereas \cite{O12} normalized $\dot M_W$ by the instantaneous surface density, which in our case would decrease in time.} to \cite{O12} (Fig.~4). This demonstrates a plausible causal transition between MRI-driven outflow solutions described here and the solutions found by \cite{O12} in the MRI stable regime.

\subsection{Comparison with a global outflow solution}
 \begin{figure*}
   \centering
   \includegraphics[width=0.44\linewidth]{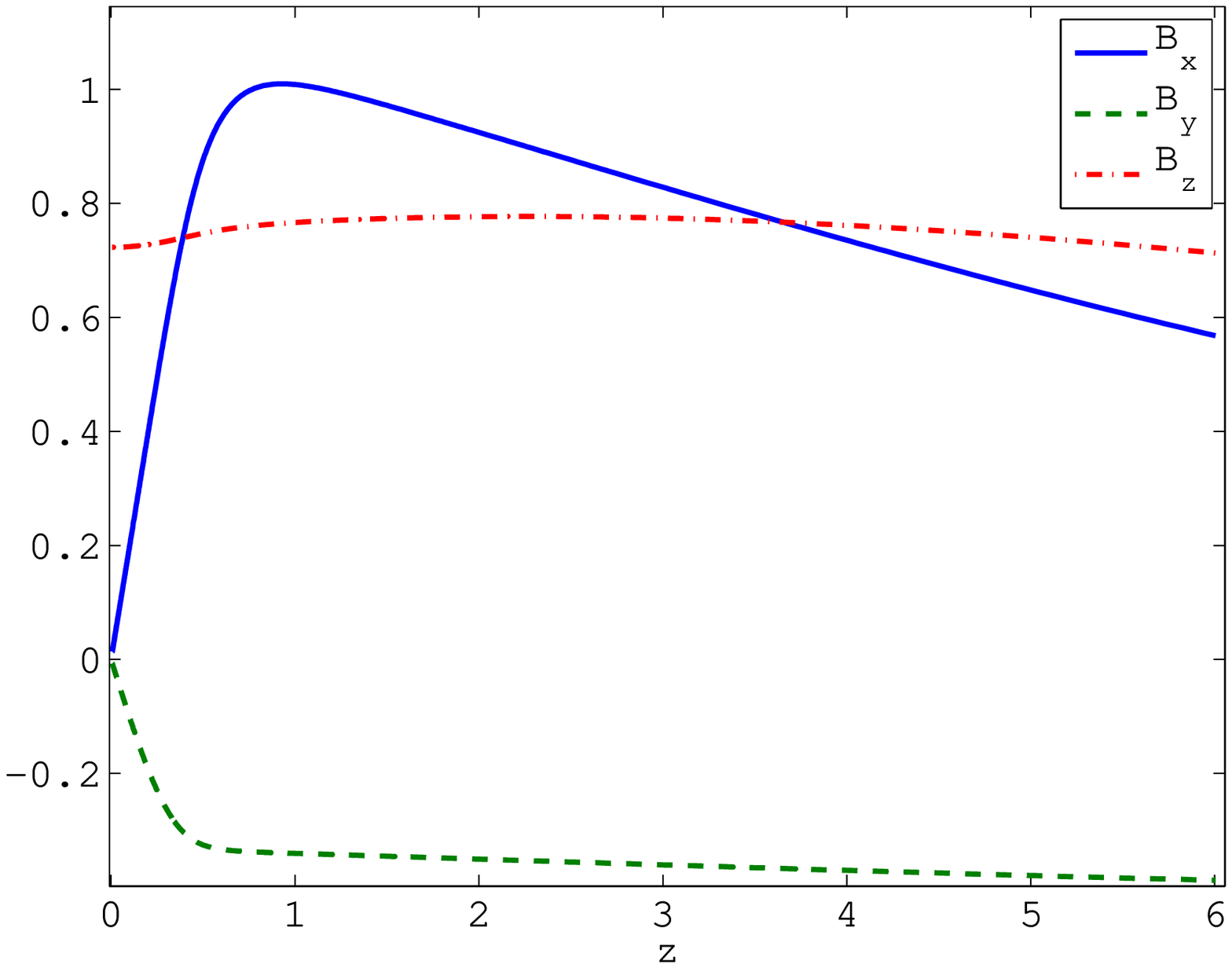}
   \includegraphics[width=0.44\linewidth]{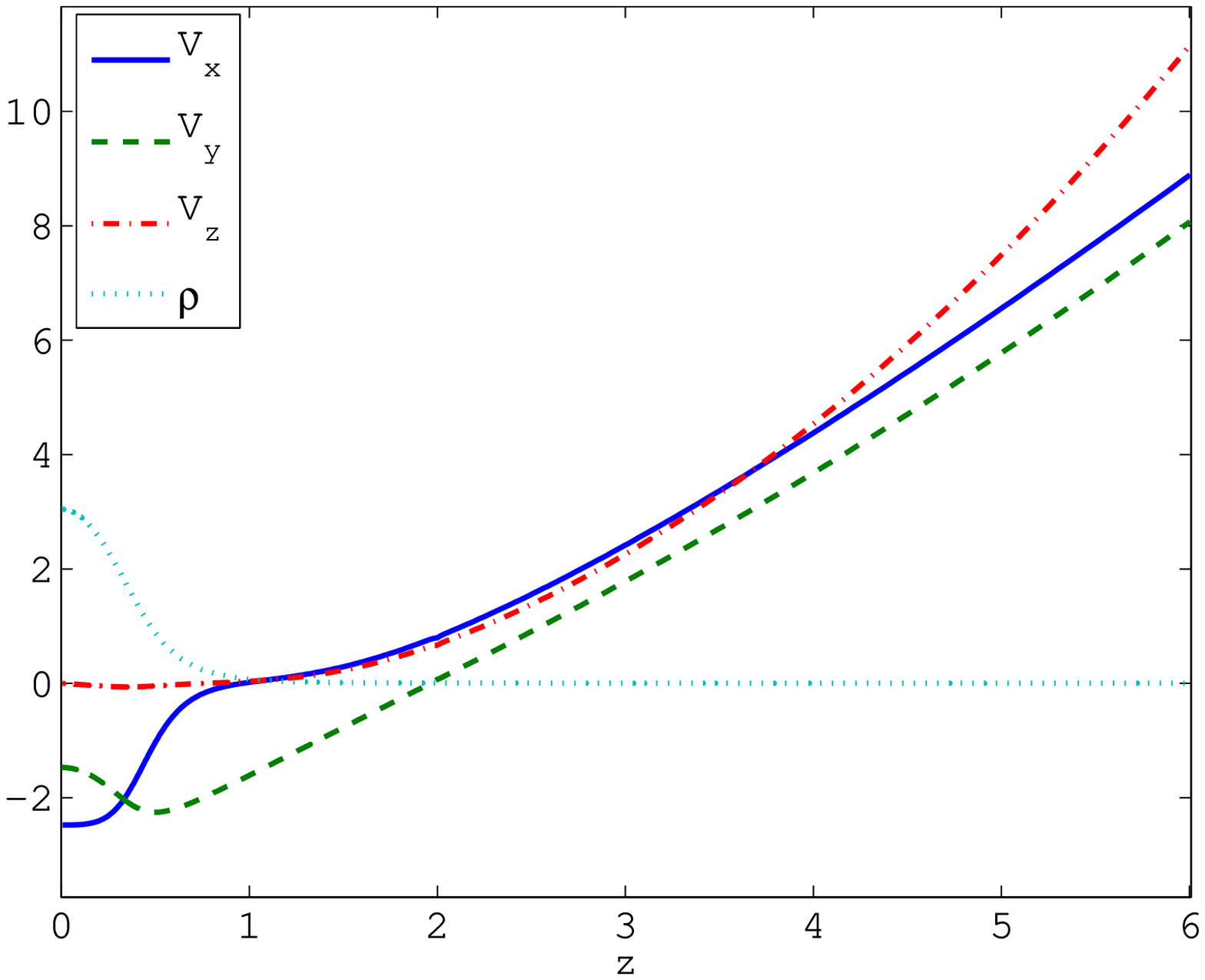}
   
   \caption{Global solution computed following \cite{FP95} in the $z=0\dots 6H$ region (see text). Colours and units identical to Fig.~\ref{Fig:profiles}. }
              \label{Fig:globsol}%
    \end{figure*}
Because the shearing box model is subject to several restrictive hypothesis, it is useful to compare the MRI driven outflows found here with global steady self-similar solutions of disc winds. The idea here is not to have a quantitive agreement between these solutions since they are computed in very different different configurations. Instead, we look for \emph{qualitative similarities} which could indicate that the physical processes at work are similar. To this end, we have used the numerical procedure described by \cite{FP95} and \cite{CF00} to compute an isothermal ``cold'' solution with the following parameters\footnote{The precise definition of these parameters can be found in \cite{CF00}. Note that the definition of the $\mu$ parameter used in these solutions differs slightly from the $\mu$ used in this paper.}: $\epsilon =0.1$ ; $\alpha_m=1.0$ ; $Pm=0$ ; $\mu=0.7268$ ; $\xi=9.92\times 10^{-3}$. This last parameter is connected to the radial dependence of the accretion rate through the relation $\dot M_a(r)\propto r^{\xi}$. Although this solution depends explicitly on $r$ and $z$ through a self-similar \emph{ansatz}, the radial dependence is much weaker than the vertical one up to a few scale heights. Therefore, we only show the $z$ dependence of the resulting solution in Fig.~\ref{Fig:globsol}. We have zoomed on the region from $z=0\dots 6H$ to allow a comparison with shearing box solutions (Fig.~\ref{Fig:profiles}), although the actual global solution extends up to $25H$ (fast magnetosonic point).

The direct comparison between this global solution and the shearing box solutions shows several differences: the velocities, and in particular $v_z$, are larger in the global solution. On the other hand, the horizontal magnetic field strength is weaker compared to the shearing box solution. However, we recover most of the physical properties found in global solutions in the shearing box solution: a strong magnetic shear close to the midplane accompanied by a magnetically compressed disc (compared to a purely hydrostatic equilibrium) and a strong accretion flow in the disc midplane ($v_x\sim-2c$).

It should be noted that the global solution presented here has a much weaker mass loss rate ($\rho v_z\sim 5\times 10^{-3}$) compared to the shearing box solution ($\rho v_z\simeq 0.1$). This can be partly due to the stronger mean vertical field required by the global solution ($\mu\sim 0.24$), although this difference is not enough to explain entirely this discrepancy (Fig.~\ref{fig:Mw_nomass}). The fact that the mass loading is much smaller in the global solution explains the faster escape velocity found in this solution.

\section{Three dimensional solution and stability\label{3Dsolutions}}
\subsection{Outflow evolution in 3D}
In order to investigate the 3D evolution of the 1D outflow desrcribed in \S{\ref{sec:steady_flow}, we have perform the outflow simulations in 3D. The initial condition consists of the 1D initial perturbation described in \S\ref{sec:1Dout} plus a random 3D noise added to $v_x$ with $\mathrm{max}(|v_x|)=0.02$. We show in Fig.~\ref{Fig:3Devol} the temporal evolution of simulation \textsc{3DRef} for $\mu=8\times 10^{-2}$. These figures represent the evolution of the density, poloidal magnetic field inclination and vertical velocity averaged in the $(x,y)$ plane in a $(z,t)$ diagram. We first observe the presence of a strong outburst ($t\sim 8$) followed by a rather steady state during which the flow does not evolve rapidly. This state corresponds to the $1D$ solution described in \S\ref{sec:1Dout} and is essentially a 1D outflow solution. However, after some time, this 1D outflow goes unstable ($t\sim 40$) and rapid variations in all quantities are observed. Surprisingly, the global structure of the 1D outflow is conserved: on time and horizontal averages, the typical $v_z$ vertical and poloidal inclination profiles are consistent with the 1D steady solution (Fig.~\ref{Fig:profiles}). We therefore produce a turbulent outflow driven by the \cite{BP82} magnetocentrifugal effect.

We show in Fig.~\ref{Fig:volrender} snapshots of the same simulation taken at $t=20$, $t=40$ and $t=270$. At $t=20$, we confirm the 1D nature of the solution: no dependence can be seen in $x$ or $y$. This  also shows that the outflow is produced before parasitic modes get significantly excited. At $t=40$ we see the development of a ``secondary'' instability of the outflow solution which produces a $x$-dependent structure. The physical processes responsible for this instability will be discussed in \S\ref{sec:3Dstability}. When the ``secondary'' modes reach a significant amplitude, the structures break down into non axisymmetric turbulent motions in which the main outflow properties are maintained (field line inclination, outflow speed). A typical example of such a state is shown at $t=270$.  
 \begin{figure*}
   \centering
   \includegraphics[width=0.99\linewidth]{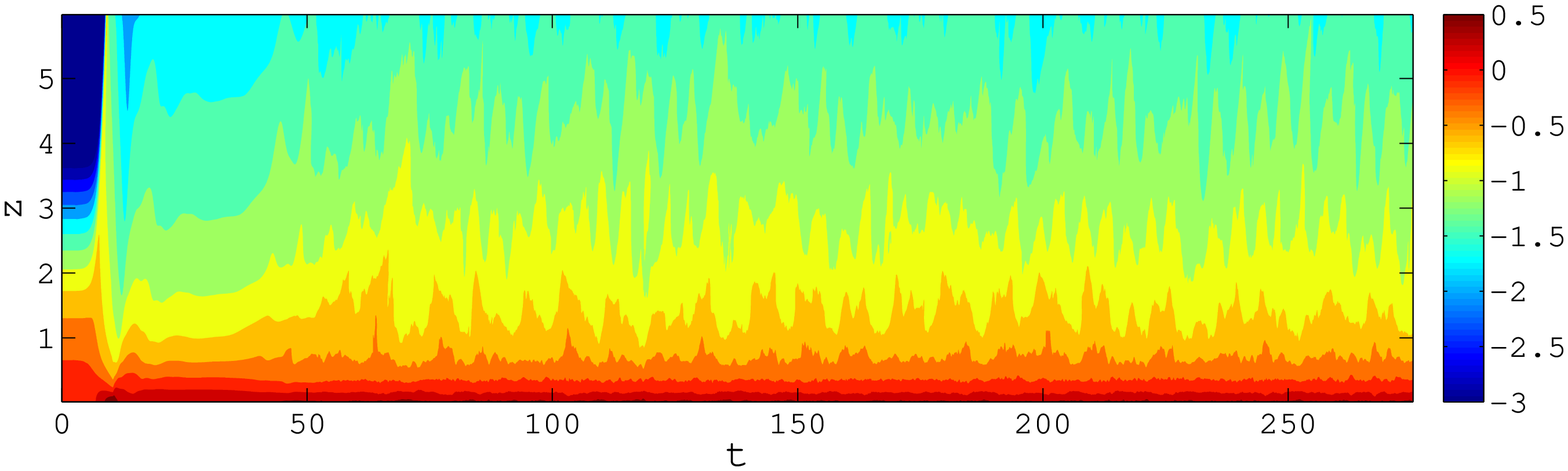}
   \includegraphics[width=0.99\linewidth]{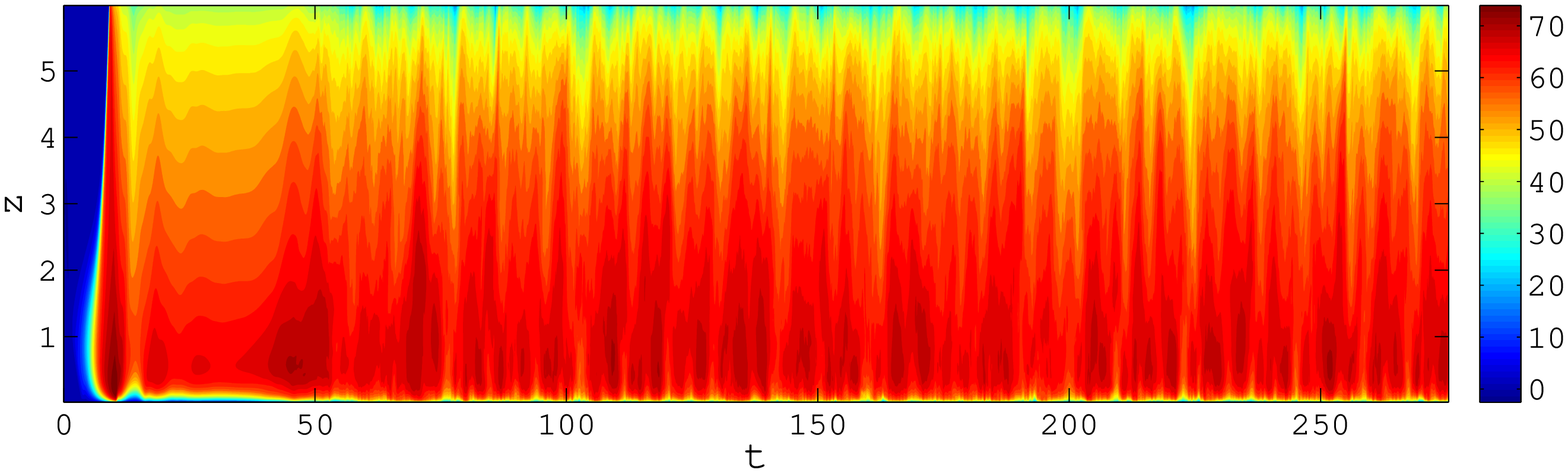}
   \includegraphics[width=0.99\linewidth]{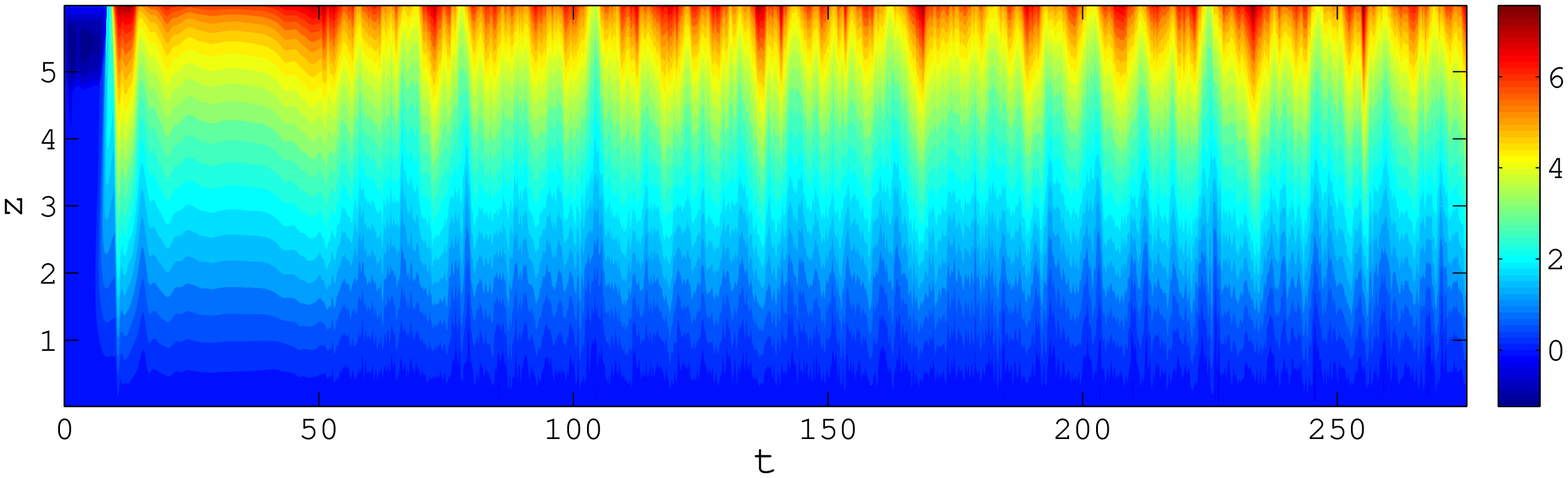}
   
   \caption{Spacetime diagram of horizontally averaged quantities in a $\mu=8\times10^{-2}$ simulation. Top: $\log_{10}(\rho)$, middle: poloidal magnetic field line inclination with respect to the vertical axis(in${\,}^\circ$), bottom: vertical velocity. }
              \label{Fig:3Devol}%
    \end{figure*}

 \begin{figure}
   \centering
   \includegraphics[width=0.99\linewidth]{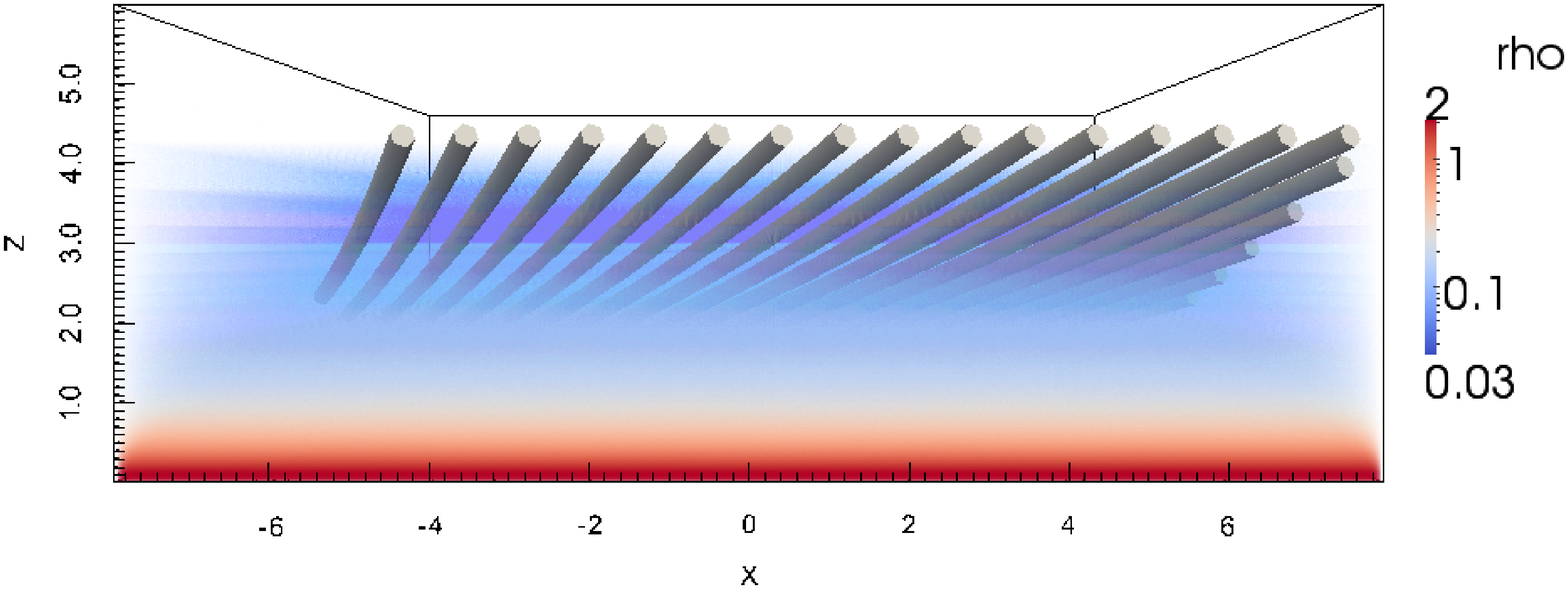}
   \includegraphics[width=0.99\linewidth]{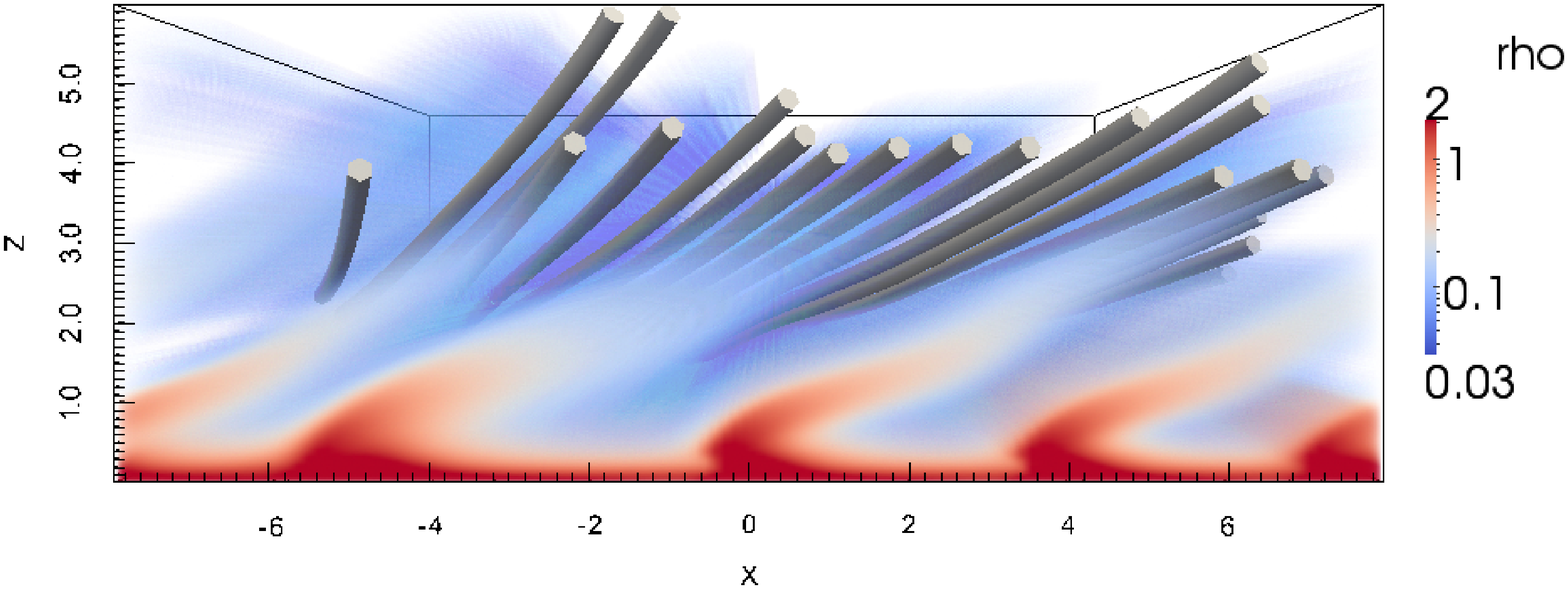}
   \includegraphics[width=0.99\linewidth]{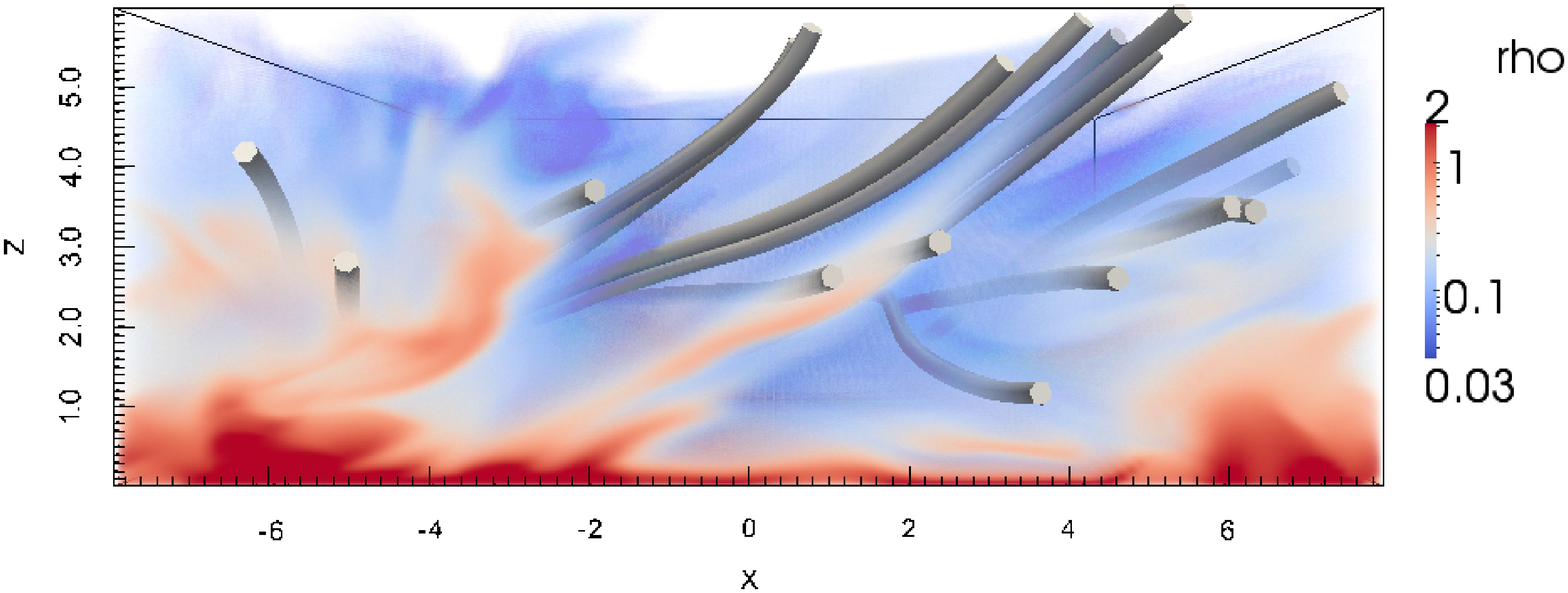}
   
   \caption{Snapshots of the $\mu=8\times 10^{-2}$ solution. Logarithm of the density is represented in coloured volume rendering whereas field lines are represented as tubes. From top to bottom: $t=20$ ; $t=40$ ; $t=270$. $x$ axis is horizontal and $z$ axis is vertical.}
              \label{Fig:volrender}%
    \end{figure}

%%%%%%%%%%%%%%%%%%%%%%%%%%%%%
% Stability analysis %
%%%%%%%%%%%%%%%%%%%%%%%%%%%%%%%
\subsection{Outflow solution stability\label{sec:3Dstability}}
\begin{figure}
   \centering
   \includegraphics[width=1.0\linewidth]{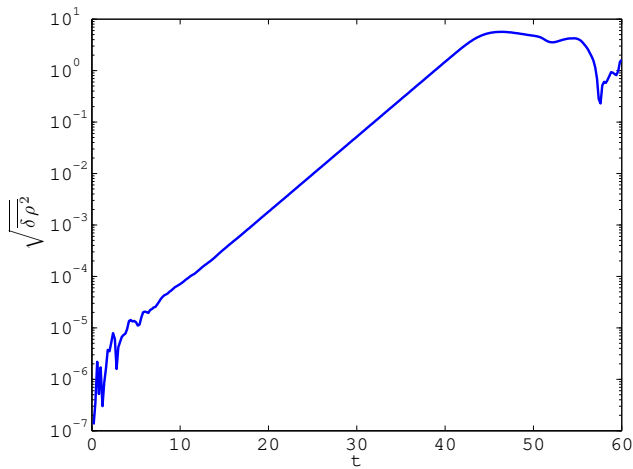}
   
   \caption{RMS amplitude of the fluctuations $\delta \rho$ in run \textsc{3DLin}. We observe a linear growth phase with $\gamma=0.33\Omega$.}
              \label{fig:growth_rate3D}%
              
 \end{figure}
 
\begin{figure}
   \centering
   \includegraphics[width=1.0\linewidth]{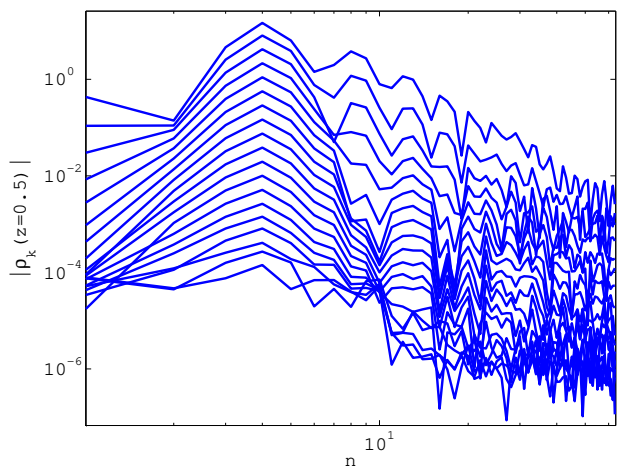}
   
   \caption{Amplitude of the Fourier modes $|\rho_k(z=0.5)|$ during the linear phase of the outflow instability. Mode decomposition are plotted every $\Delta t=2$ from $t=0$ to $t=40$.  The instability is dominated by the $n=4$ mode (see text).}
              \label{fig:mode_amplitude}%
              
 \end{figure}
 
 \begin{figure}
   \centering
   \includegraphics[width=1.0\linewidth]{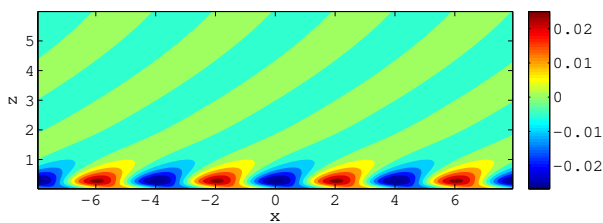}
   
   \caption{Density fluctuations corresponding to the $n=4$ eigenmode.}
              \label{fig:eigenmodes}%
              
 \end{figure}
 
\begin{figure}
   \centering
   \includegraphics[width=1.0\linewidth]{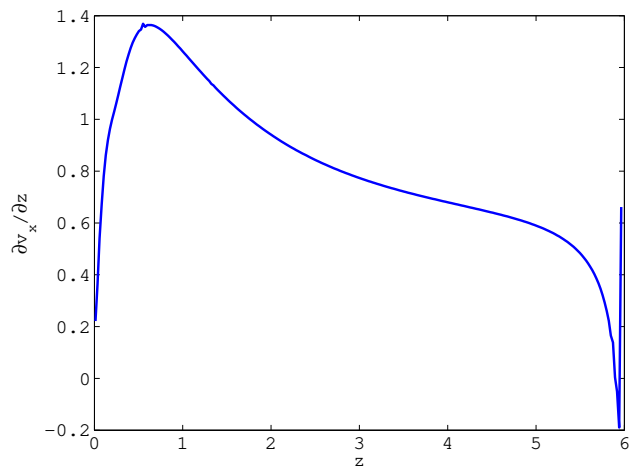}
   
   \caption{$y$ component of the vorticity in the stationary outflow solution.}
              \label{fig:vorticity}%
              
 \end{figure}
 
 \begin{figure}
   \centering
   \includegraphics[width=1.0\linewidth]{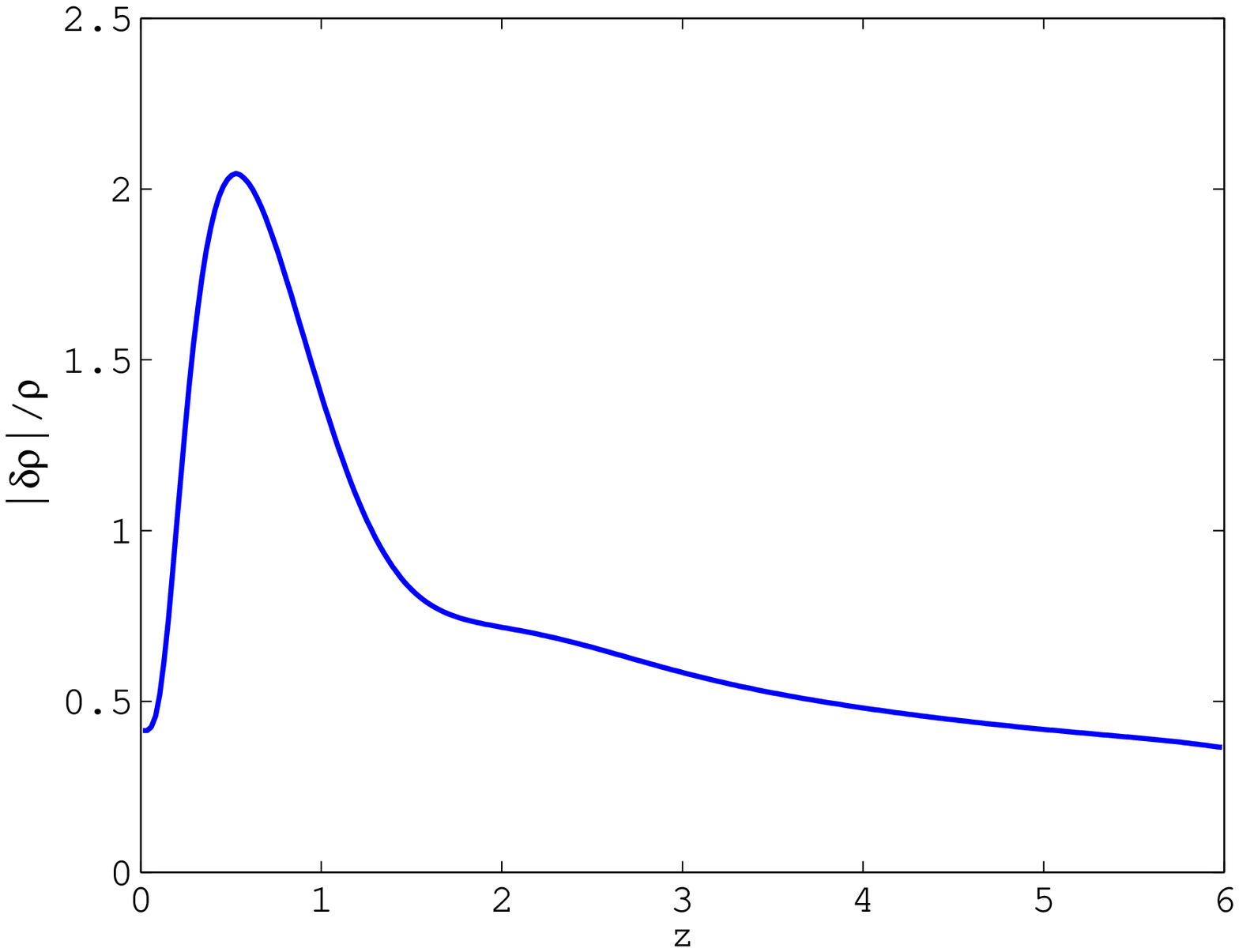}
   
   \caption{Eigenmode $n=4$ vertical density profile}
              \label{fig:eigenmode_profile}%
              
 \end{figure}

We have shown above that the outflow solution is unstable to 3D perturbation, resulting in a ``turbulent outflow'' configuration. Several authors have discussed the possibility of having such an instability \citep{LPP94b,LS95,CS02} although the applicability of these stability analyses to all outflow solutions is still uncertain \citep{KW96,KO04}. 

In order to analyse the instability observed in our outflow solutions, we have performed a simulation (\textsc{3DLin}) starting from the 1D solution corresponding to the final state of run \textsc{1DRef} to which a small-amplitude ($10^{-3}$) 3D white noise was added. Since the growth phase of the instability implies only $x$-dependent modes, we use a Fourier decomposition in the $x$ direction to characterise growing modes:
\begin{eqnarray}
\rho&=&\rho_0(z,t)+\delta\rho(x,z,t)\\
\delta\rho(x,z,t)&=&\Re\Big[ \sum_k\rho_k(z,t)\exp(ikx)\Big]
\end{eqnarray}
and similarly for $\bm{v}$ and $\bm{B}$. In this expansion, we have assumed the instability was growing on the top of the 1D solution $\rho_0(z)$ given by the final state of run \textsc{1DRef}. We first present the temporal evolution of the fluctuation $\sqrt{\langle\delta\rho^2\rangle}$ in Fig.~\ref{fig:growth_rate3D}. We find that perturbations grow exponentially with a growth rate $\gamma=0.33\Omega$  up to $t\simeq 40$ where a saturation is reached. 
To further investigate the behaviour of this instability, we present the temporal evolution of $|\rho_k(z=0.5)|$ as a function of $k$ in Fig.~\ref{fig:mode_amplitude}. As it can readily be seen, the growth is dominated by modes having $n\equiv kL_x/2\pi=4$ which is consistent with the 4 ``spots'' observed in Fig.~\ref{Fig:volrender}. The measured growth rate of this mode is $\gamma=0.33\Omega$ which explains its fast appearance in 3D simulations once an outflow has formed. However, other neighbouring modes are growing as well ($n=3;5;6$) although not as fast as the $n=4$ mode. Finally once the $n=4$ mode reaches large amplitudes ($|\rho_k|\sim1$), we note the sudden growth of the $n=8$ and $n=9$ modes which are the result of nonlinear interactions of the fastest growing modes $n=3;4;5;6$. 

In order to analyse the physics underlying the $n=4$ mode, we present in Fig.~\ref{fig:eigenmodes} the density fluctuations corresponding to that eigenmode. We first note that the density fluctuation is highly localised in $z$ around $z\sim0.5$. In addition, this eigenmode has a tail whose inclination and shape closely follow that of the poloidal magnetic field (Fig.~\ref{fig:streamlines}). The localisation of the eigenmode is rather surprising and requires some explanation. We first note that this localisation is much higher than the top of the mass injection region ($H_\mathrm{inj}=0.1$). However, comparing the relative vorticity component $\omega_y=\partial_zv_x$ of the 1D outflow (Fig.~\ref{fig:vorticity}) to the vertical profile of the eigenmode (Fig~\ref{fig:eigenmode_profile}), we find that the density perturbation is localised close to a maximum of $\omega_y$. This tends to suggest that this instability is somehow linked to the vertical $\omega_y$ profile of the outflow solution, and therefore to a kind of Kelvin-Helmholtz instability. This potential link is also consistent with the growth rate $\gamma\lesssim \mathrm{max}(\omega_y)$ and with the shape of the eigenmode and the vorticity profile around $z=0.5$.

We would like to stress that these remarks are not a proof that this outflow instability is of the Kelvin-Helmholtz type. Among the effects we did not take into account in that analysis are the magnetic field, compressibility and the presence of a large $v_z$ up in the atmosphere. However, if we assume that the source of the instability lies around $z\sim0.5$ as suggested by the eigenmodes, then time scales, length scales and phenomenology match that of a Kelvin-Helmholtz instability. To ascertain these claims, a proper linear study taking into account all of the outflow properties is required, which is well beyond the scope of this paper.

\section{Discussion}

In this paper, we have shown that large scale MRI modes which are unstable when the disc magnetisation is moderately sub-thermal spontaneously produce super-Alfv\'enic outflows. The physical mechanism behind this outflow is a Blandford \& Payne-like process where angular momentum is transferred to bending field line and then released to accelerated material. We demonstrated that this outflow is qualitatively similar to the outflow solutions found both in local boxes and in global self-similar geometry, making a clear connection between the MRI and the formation of disc winds. We have also shown that MRI outflows are unstable in 3D which could be a potential source of variability for disc winds and jets. These 3D instabilities could also be the origin of the turbulent resistivity $\alpha_m$  used in \cite{FP95} and subsequent self-similar models.

However, the picture provided here is still incomplete. We first note that the simulations we have performed here only produce super-Alfv\'enic flows, whereas real escaping outflows should be faster than the fast magnetosonic speed. This problem seems to be linked to the shearing box geometry, as several other authors have noticed this in shearing boxes (\citealt{SI09}, \citealt{FLLO12}, X.~Bai private communication). Indeed, a shearing box does not allow for the opening of magnetic field lines which is expected in realistic global geometry. We believe that such an opening is required to get super-fast flows, which are therefore beyond the scope of our shearing box model. The fact that the flow is only super-Alfv\'enic indicates that the upper boundary condition still plays a role in determining the outflow structure, and indeed it does, at least for the mass-loss rate (\S \ref{sec:upperbound}).

We also emphasize that the  shearing box model possesses a horizontal symmetry by the transformation $(v_x,v_y)\rightarrow(-v_x,-v_y)$, $(B_x,B_y)\rightarrow(-B_x,-B_y)$ and $(x,y)\rightarrow (-x,-y)$.  This symmetry indicates that locally, there is no mathematical difference between a magneto-centrifugally accelerated wind where angular momentum is transferred from the disc to the jet and a magnetically decelerated accretion column (formally a magneto-centripetal wind) where the angular momentum is transferred from material falling radially inward to the disc.  This symmetry is obviously broken once curvature terms are taken into account. The presence of this symmetry implies that shearing box simulations can spontaneously switch from one situation to the other, which is unexpected in realistic situations. Such sudden changes in the magnetic configuration were indeed observed in rare occasions in our 3D runs but also by  \cite{FLLO12}.

Finally, we should point out that the presence of a non-zero toroidal electromotive force implies that magnetic field lines are accreted. As mentioned earlier, this situation is rather unrealistic in global geometry, although it is allowed in shearing boxes. More realistic configurations, probably including a sort of resistivity (either effective or molecular) are required to compensate for this effect.

All of these points indicate that shearing boxes are not very well suited to study globally the outflows produced by MRI turbulent accretion discs. In particular, little can learned regarding outflow mass loss rate and velocities. We note however that our solutions can be qualitatively compared to global solution and several properties are recovered by the local model. Moreover, 3D instabilities can be studied much more easily in boxes than in global geometry where computational costs increase very rapidly.

Overall, our results tend to indicate a paradigm shift: up to now, the MRI driving ``viscous'' discs and disc winds at the origin of jets have often been considered as separate processes. Here we show that these two processes are actually intrinsically connected: \emph{outflows are a logical consequence of the MRI in strongly magnetised discs}. Obviously, the next question is to understand what is driving the disc magnetisation. This could be due to local dynamos, large scale field redistribution through advection and diffusion, etc. To identify these processes, shearing boxes are clearly insufficient and global models, including both large-scale magnetic fields and turbulence, will be required.

\section{Appendix}

\begin{table*}
\centering
% Name
\begin{tabular}{|c|c|c|c|c|c|c|}
\hline
Run & Resolution                    & Box size                                     & Mass injection & Smoothed potential &  Boundary condition   & Outflow\\
\hline 
\hline
\textsc{1DRef}              &  $1\times1\times256$   &   $1.0\times 1.0\times 6.0$         &  Yes		     & No			             &  $B_x(z_B)=0$             &  Yes   \\	% test40 (osug)
\hline
\textsc{1Dz4}               &  $1\times1\times170$   &   $1.0\times 1.0\times 4.0$         &  Yes		     & Yes			             &  $B_x(z_B)=0$             &  Yes   \\	% test48 (osug)
\hline
\textsc{1Dz6}               &  $1\times1\times256$   &   $1.0\times 1.0\times 6.0$         &  Yes		     & Yes			             &  $B_x(z_B)=0$             &  Yes   \\	% test53 (osug)
\hline
\textsc{1Dz8}               &  $1\times1\times340$   &   $1.0\times 1.0\times 8.0$         &  Yes		     & Yes			             &  $B_x(z_B)=0$             &  Yes   \\	% test49 (osug)
\hline
\textsc{1Dz12}	       &  $1\times1\times512$   &   $1.0\times 1.0\times 12.0$         &  Yes		     & Yes			             &  $B_x(z_B)=0$             &   Yes  \\	% test50 (osug)
\hline
\textsc{1Dz16}	       &  $1\times1\times684$   &   $1.0\times 1.0\times 16.0$         &  Yes		     & Yes			             &  $B_x(z_B)=0$             &   Yes  \\	% test51 (osug)
\hline
\textsc{1Dz20}	       &  $1\times1\times856$   &   $1.0\times 1.0\times 20.0$         &  Yes		     & Yes			             &  $B_x(z_B)=0$             &   Yes  \\	% test52 (osug)
\hline
\textsc{1DZG}	       &  $1\times1\times256$   &   $1.0\times 1.0\times 6.0$         &  Yes		     & No			             &  $\partial_zB_x(z_B)=0$             &  No   \\	% test43 (osug)
\hline
\textsc{1DInc}	       &  $1\times1\times256$   &   $1.0\times 1.0\times 6.0$         &  Yes		     & No			             &   $\left\{ \begin{array}{ll} B_x(z_B)=B_z(z_B)\\B_y(z_B)=0\end{array}\right.$             &  Yes   \\	% test58 (osug)
\hline
\textsc{1DNoMass}      &  $1\times1\times256$   &   $1.0\times 1.0\times 6.0$         &  No		     & No			             &  $B_x(z_B)=0$             &  No${\,}^a$   \\	% test54 (osug)
\hline

\textsc{1DNoMassInc} &  $1\times1\times256$   &   $1.0\times 1.0\times 6.0$         &  No		     & No			             &  $\left\{ \begin{array}{ll} B_x(z_B)=B_z(z_B)\\B_y(z_B)=0\end{array}\right.$             &  Yes   \\	% test56 (osug)
\hline
\hline
\textsc{3DRef}		& $128\times128\times256$   &   $16.0\times 16.0\times 6.0$         &  Yes		     & No			             &  $B_x(z_B)=0$             &  Yes   \\	% test37_3D_beta10 (IDRIS)
\hline
\textsc{3DLin}${\,}^b$		& $128\times128\times256$   &   $16.0\times 16.0\times 6.0$         &  Yes		     & No			             &  $B_x(z_B)=0$             &  Yes   \\	% test37-bis (R2D2)
\hline
\end{tabular}

\caption{\label{tab:runs}List of the simulations discussed in this paper. \newline ${\,}^a$: The outflow disappears when the Alfv\'en point gets out of the simulation box (\S\ref{sec:mag_dependency})   \newline ${\,}^b$: The initial condition for this simulation corresponds to the final state of \textsc{1DRef} to which small amplitude 3D white noise was added (\S\ref{sec:3Dstability})}
\end{table*}

\begin{acknowledgements}
      GL thanks S\'ebastien Fromang for his comments and suggestions on the initial version of this manuscript. This work was granted access to the HPC resources of IDRIS under allocation x2012042231 made by GENCI (Grand Equipement National de Calcul Intensif). Some of the computations presented in this paper were performed using the CIMENT infrastructure (https://ciment.ujf-grenoble.fr), which is supported by the Rhone-Alpes region (GRANT CPER07\_13 CIRA: http://www.ci-ra.org).

      \end{acknowledgements}

\bibliographystyle{aa}
\bibliography{glesur}

\begin{thebibliography}{42}
\expandafter\ifx\csname natexlab\endcsname\relax\def\natexlab#1{#1}\fi

\bibitem[{{Armitage}(2011)}]{A11}
{Armitage}, P.~J. 2011, \araa, 49, 195

\bibitem[{{Balbus}(2003)}]{B03}
{Balbus}, S.~A. 2003, \araa, 41, 555

\bibitem[{{Balbus} \& {Hawley}(1991)}]{BH91a}
{Balbus}, S.~A. \& {Hawley}, J.~F. 1991, \apj, 376, 214

\bibitem[{{Blandford} \& {Payne}(1982)}]{BP82}
{Blandford}, R.~D. \& {Payne}, D.~G. 1982, \mnras, 199, 883

\bibitem[{{Cao} \& {Spruit}(2002)}]{CS02}
{Cao}, X. \& {Spruit}, H.~C. 2002, \aap, 385, 289

\bibitem[{{Casse} \& {Ferreira}(2000)}]{CF00}
{Casse}, F. \& {Ferreira}, J. 2000, \aap, 353, 1115

\bibitem[{{Chandrasekhar}(1956)}]{C56}
{Chandrasekhar}, S. 1956, \apj, 124, 232

\bibitem[{{Evans} \& {Hawley}(1988)}]{EH88}
{Evans}, C.~R. \& {Hawley}, J.~F. 1988, \apj, 332, 659

\bibitem[{{Ferreira} \& {Pelletier}(1993{\natexlab{a}})}]{FP93}
{Ferreira}, J. \& {Pelletier}, G. 1993{\natexlab{a}}, \aap, 276, 625

\bibitem[{{Ferreira} \& {Pelletier}(1993{\natexlab{b}})}]{FP93b}
{Ferreira}, J. \& {Pelletier}, G. 1993{\natexlab{b}}, \aap, 276, 637

\bibitem[{{Ferreira} \& {Pelletier}(1995)}]{FP95}
{Ferreira}, J. \& {Pelletier}, G. 1995, \aap, 295, 807

\bibitem[{{Flock} {et~al.}(2011){Flock}, {Dzyurkevich}, {Klahr}, {Turner}, \&
  {Henning}}]{FD11}
{Flock}, M., {Dzyurkevich}, N., {Klahr}, H., {Turner}, N.~J., \& {Henning}, T.
  2011, \apj, 735, 122

\bibitem[{{Fromang} {et~al.}(2012){Fromang}, {Latter}, {Lesur}, \&
  {Ogilvie}}]{FLLO12}
{Fromang}, S., {Latter}, H., {Lesur}, G., \& {Ogilvie}, G.~I. 2012, submitted
  to\aap

\bibitem[{{Fromang} {et~al.}(2011){Fromang}, {Lyra}, \& {Masset}}]{FL11}
{Fromang}, S., {Lyra}, W., \& {Masset}, F. 2011, \aap, 534, A107

\bibitem[{{Goodman} \& {Xu}(1994)}]{GX94}
{Goodman}, J. \& {Xu}, G. 1994, \apj, 432, 213

\bibitem[{{Hawley}(2000)}]{H00}
{Hawley}, J.~F. 2000, \apj, 528, 462

\bibitem[{{Hawley} {et~al.}(1995){Hawley}, {Gammie}, \& {Balbus}}]{HGB95}
{Hawley}, J.~F., {Gammie}, C.~F., \& {Balbus}, S.~A. 1995, \apj, 440, 742

\bibitem[{{K{\"o}nigl}(2004)}]{KO04}
{K{\"o}nigl}, A. 2004, \apj, 617, 1267

\bibitem[{{K{\"o}nigl} {et~al.}(2010){K{\"o}nigl}, {Salmeron}, \&
  {Wardle}}]{KSW10}
{K{\"o}nigl}, A., {Salmeron}, R., \& {Wardle}, M. 2010, \mnras, 401, 479

\bibitem[{{K{\"o}nigl} \& {Wardle}(1996)}]{KW96}
{K{\"o}nigl}, A. \& {Wardle}, M. 1996, \mnras, 279, L61

\bibitem[{Latter {et~al.}(2010)Latter, Fromang, \& Gressel}]{LF10}
Latter, H.~N., Fromang, S., \& Gressel, O. 2010, MNRAS, 406, 848

\bibitem[{{Latter} {et~al.}(2009){Latter}, {Lesaffre}, \& {Balbus}}]{LLB09}
{Latter}, H.~N., {Lesaffre}, P., \& {Balbus}, S.~A. 2009, \mnras, 394, 715

\bibitem[{{Lemaster} \& {Stone}(2009)}]{LS09}
{Lemaster}, M.~N. \& {Stone}, J.~M. 2009, \apj, 691, 1092

\bibitem[{{Lesur} \& {Longaretti}(2011)}]{LL11}
{Lesur}, G. \& {Longaretti}, P.-Y. 2011, \aap, 528, A17

\bibitem[{{Longaretti} \& {Lesur}(2010)}]{LL10}
{Longaretti}, P. \& {Lesur}, G. 2010, \aap, 516, A51+

\bibitem[{{Lubow} {et~al.}(1994){Lubow}, {Papaloizou}, \& {Pringle}}]{LPP94b}
{Lubow}, S.~H., {Papaloizou}, J.~C.~B., \& {Pringle}, J.~E. 1994, \mnras, 268,
  1010

\bibitem[{{Lubow} \& {Spruit}(1995)}]{LS95}
{Lubow}, S.~H. \& {Spruit}, H.~C. 1995, \apj, 445, 337

\bibitem[{{Lynden-Bell} \& {Pringle}(1974)}]{LP74}
{Lynden-Bell}, D. \& {Pringle}, J.~E. 1974, \mnras, 168, 603

\bibitem[{{Mestel}(1961)}]{M61}
{Mestel}, L. 1961, \mnras, 122, 473

\bibitem[{{Mignone} {et~al.}(2007){Mignone}, {Bodo}, {Massaglia}, {Matsakos},
  {Tesileanu}, {Zanni}, \& {Ferrari}}]{MBM07}
{Mignone}, A., {Bodo}, G., {Massaglia}, S., {et~al.} 2007, \apjs, 170, 228

\bibitem[{{Ogilvie}(2012)}]{O12}
{Ogilvie}, G.~I. 2012, \mnras, 423, 1318

\bibitem[{{Pelletier} \& {Pudritz}(1992)}]{PP92}
{Pelletier}, G. \& {Pudritz}, R.~E. 1992, \apj, 394, 117

\bibitem[{{Pessah}(2010)}]{P10}
{Pessah}, M.~E. 2010, \apj, 716, 1012

\bibitem[{{Pessah} \& {Goodman}(2009)}]{PG09}
{Pessah}, M.~E. \& {Goodman}, J. 2009, \apjl, 698, L72

\bibitem[{{Pudritz} \& {Norman}(1983)}]{PN83}
{Pudritz}, R.~E. \& {Norman}, C.~A. 1983, \apj, 274, 677

\bibitem[{{Regev} \& {Umurhan}(2008)}]{RU08}
{Regev}, O. \& {Umurhan}, O.~M. 2008, \aap, 481, 21

\bibitem[{{Shakura} \& {Sunyaev}(1973)}]{SS73}
{Shakura}, N.~I. \& {Sunyaev}, R.~A. 1973, \aap, 24, 337

\bibitem[{{Spruit}(1996)}]{S96}
{Spruit}, H.~C. 1996, {Magnetohydrodynamic winds and jets from accretion
  disks}, Vol. 477 (Klewer), 249

\bibitem[{{Stone} {et~al.}(1996){Stone}, {Hawley}, {Gammie}, \&
  {Balbus}}]{SHGB96}
{Stone}, J.~M., {Hawley}, J.~F., {Gammie}, C.~F., \& {Balbus}, S.~A. 1996,
  \apj, 463, 656

\bibitem[{{Suzuki} \& {Inutsuka}(2009)}]{SI09}
{Suzuki}, T.~K. \& {Inutsuka}, S.-i. 2009, \apjl, 691, L49

\bibitem[{{Wardle} \& {K{\"o}nigl}(1993)}]{WK93}
{Wardle}, M. \& {K{\"o}nigl}, A. 1993, \apj, 410, 218

\bibitem[{{Zanni} {et~al.}(2007){Zanni}, {Ferrari}, {Rosner}, {Bodo}, \&
  {Massaglia}}]{ZF07}
{Zanni}, C., {Ferrari}, A., {Rosner}, R., {Bodo}, G., \& {Massaglia}, S. 2007,
  \aap, 469, 811

\end{thebibliography}

\end{document}